\documentclass[superscriptaddress,preprint,preprintnumbers,nofootinbib,amsmath,amssymb,aps,prd]{revtex4-2}

\usepackage{graphicx}
\usepackage{dcolumn}
\usepackage{bm}
\usepackage{array} 
\newcolumntype{M}[1]{>{\centering\arraybackslash}m{#1}}
\usepackage{endnotes} 
\usepackage{adjustbox}

\usepackage{amssymb}    
\usepackage{color}         
\usepackage{graphicx}     
\usepackage{mathrsfs} 
\usepackage{hyperref} 
\hypersetup{colorlinks=true,linkcolor=blue,citecolor=blue}
\usepackage{ upgreek } 
\usepackage{color,xcolor,fancybox,epsf,rotating,colordvi}
\usepackage{booktabs}
\usepackage{multirow}

\usepackage{soul}
\setstcolor{red}

\usepackage{cancel}

\begin{document}


\title{Linking a 152 GeV diphoton excess to dark matter and gravitational waves in the singlet-extended 2HDM}

\author{Yabo Dong}
\affiliation{School of Physics and Electronics, Henan University, Kaifeng 475004, China}

\author{Jun Fan}
\affiliation{School of Physics and Electronics, Henan University, Kaifeng 475004, China}

\author{Kun Wang}
\email[Corresponding author:]{kwang@usst.edu.cn} 
\affiliation{College of Science, University of Shanghai for Science and Technology, Shanghai 200093, China}

\author{Haijun Yang}
\affiliation{State Key Laboratory of Dark Matter Physics, Key Laboratory for Particle Astrophysics and Cosmology (MOE), Shanghai Key Laboratory for Particle Physics and Cosmology (SKLPPC),  School of Physics and Astronomy \mbox{\normalfont\&} Tsung-Dao Lee Institute, Shanghai Jiao Tong University, Shanghai 200240, China}

\author{Jingya Zhu}
\email[Corresponding author:]{zhujy@henu.edu.cn} 
\affiliation{School of Physics and Electronics, Henan University, Kaifeng 475004, China}


\date[]{Dated: October 28, 2026}

\begin{abstract}
We investigate the diphoton excesses near $152~\mathrm{GeV}$ accompanied by missing transverse momentum (MET) in the Type-I two-Higgs-doublet model extended by a complex scalar singlet, whose imaginary component serves as a pseudo-Nambu--Goldstone dark matter (DM) candidate $\chi$.
The signal arises from the cascade process $gg\to h_3\to h_2h_2\to\gamma\gamma\chi\chi$, where $h_2$ is the scalar resonance near $152~\mathrm{GeV}$ and $h_3$ is a heavier CP-even scalar.
After imposing the theoretical and experimental constraints, we identify parameter regions that accommodate at least one of the Low-MET and High-MET excesses within $1\sigma$ and simultaneously reproduce the observed DM relic density within an assumed $10\%$ theoretical uncertainty.
A subset of these samples supports first-order electroweak phase transitions with substantial supercooling.
For a three-year observation, selected samples yield gravitational-wave signals with signal-to-noise ratios (SNRs) above $10$ at LISA, TianQin, DECIGO, and BBO.
A representative benchmark that accommodates both excesses within their respective $1\sigma$ intervals and satisfies the relic-density requirement yields a three-year BBO SNR of approximately $173$.
\end{abstract}

\maketitle
\newpage



\section{Introduction}
\label{sec:intro}

The discovery of a 125~GeV Standard Model (SM)-like Higgs boson by the CMS and ATLAS collaborations in 2012 completed the particle content of the SM~\cite{CMS:2012qbp,ATLAS:2012yve}. 
Although subsequent measurements of its properties have remained consistent with the SM predictions~\cite{CMS:2022dwd,ATLAS:2022vkf}, several fundamental questions remain unanswered, including the nature of dark matter (DM)~\cite{Bertone:2004pz,Feng:2010gw} and the origin of the observed matter--antimatter asymmetry. 
These open questions provide compelling motivation for physics beyond the SM (BSM). 
In particular, extensions of the scalar sector offer a well-motivated approach to addressing some of these shortcomings while retaining a Higgs boson compatible with the observed 125~GeV state~\cite{Branco:2011iw}. 
Additional scalar particles can act as mediators connecting the dark sector to the SM or serve as DM candidates themselves.
They can also modify the finite-temperature scalar potential and allow a first-order electroweak phase transition, providing the departure from thermal equilibrium needed for electroweak baryogenesis when accompanied by sufficient CP violation~\cite{Sakharov:1967dj,Dorsch:2016nrg}.
Extended Higgs sectors therefore have implications for both particle phenomenology and the evolution of the early Universe~\cite{Mazumdar:2018dfl,Kannike:2019wsn,Alanne:2020jwx,Beniwal:2017eik,Chen:2019ebq}.

Beyond these theoretical motivations, phenomenological reanalyses of LHC data, including the sideband regions of SM Higgs searches, have suggested possible excesses near $152~\mathrm{GeV}$ in associated diphoton final states, commonly denoted by $\gamma\gamma+X$~\cite{Crivellin:2021ubm,Bhattacharya:2023lmu,Bhattacharya:2025rfr}.
Depending on the signal category, $X$ can include missing transverse momentum, leptons, jets, or combinations of these objects.
We use MET to refer to the magnitude of the missing transverse momentum, denoted by $E_T^{\mathrm{miss}}$.
Using the ATLAS data~\cite{ATLAS:2021jbf}, Ref.~\cite{Bhattacharya:2025rfr} estimates the significances of the $\gamma\gamma+\mathrm{MET}$ excesses at $2.8\sigma$ and $2.5\sigma$ in the Low-MET and High-MET categories, respectively.
This channel is particularly interesting because of its possible connection to an invisible sector containing DM.
These excesses motivate investigating their possible origin within extensions of the Higgs sector.

Within the two-Higgs-doublet model (2HDM) framework, Ref.~\cite{Banik:2024ftv} considers the Drell--Yan process $pp\to W^*\to\phi H^\pm$, where the neutral state $\phi$ decays into two photons and the charged Higgs decays into fermionic final states such as $\tau\nu$ and $t^{(*)}b$. 
Another interpretation invokes the cascade $gg\to H\to\phi\phi^{(*)}$ through a heavier scalar $H$, with the daughter-scalar decays generating various associated final states, including the $\gamma\gamma b\bar{b}$ channel proposed as a complementary probe~\cite{Crivellin:2021ubm}.
Ref.~\cite{Chen:2025vtg} investigates an interpretation in an extended Georgi--Machacek model, where mild custodial-symmetry breaking modifies the relative couplings to $WW$ and $ZZ$, and additional charged scalars can enhance the diphoton decay rate. 
An important question is whether the $\gamma\gamma+\mathrm{MET}$ excesses and the full observed DM relic density can be explained simultaneously within a single framework.

A possible framework for addressing this question is the singlet-extended two-Higgs-doublet model (S2HDM), which introduces a complex scalar singlet in addition to the two Higgs doublets~\cite{Jiang:2019soj,Biekotter:2021ovi,Biekotter:2022bxp}.
The scalar potential possesses a softly broken global $U(1)$ symmetry.
After the singlet acquires a nonzero real vacuum expectation value (VEV), its angular mode corresponds to a pseudo-Nambu--Goldstone boson (pNGB), $\chi$, and the soft symmetry-breaking term generates its mass.
An exact conjugation symmetry, $S\to S^*$, remains unbroken by the vacuum and ensures that $\chi$ is stable, allowing it to serve as a DM candidate.
The mixed CP-even scalars mediate DM annihilation into SM particles and can decay invisibly into $\chi\chi$ when kinematically allowed.
Moreover, the tree-level DM--nucleon scattering amplitude vanishes in the zero-momentum-transfer limit, suppressing direct-detection signals at this order without requiring the annihilation interactions to be negligible~\cite{Gross:2017dan,Jiang:2019soj,Karamitros:2019ewv}.
Related pNGB DM models have also been investigated through loop-level direct detection~\cite{Azevedo:2018exj,Ishiwata:2018sdi,Glaus:2020ihj}, collider probes, and global fits~\cite{Huitu:2018gbc,Arina:2019tib}.

The theoretical properties and phenomenology of the 2HDM and its singlet extensions have been extensively studied~\cite{Glashow:1976nt,Paschos:1976ay,Ginzburg:2004vp,Wang:2022yhm,Dong:2024ipo,Baum:2018zhf,Heinemeyer:2021msz,Dong:2025orv,Dong:2025exu,Biekotter:2023oen,Biekotter:2025vxl,Dutta:2023cig,Dutta:2025nmy,Cline:1996mga,Basler:2016obg,Dorsch:2017nza,Han:2020ekm,Zhang:2021alu,Chen:2013jvg,Muhlleitner:2016mzt,Ferreira:2019iqb,Biekotter:2021ysx,Dutta:2022qeq}.
In this work, we consider the CP-conserving Type-I S2HDM, whose three CP-even Higgs bosons are denoted by $h_1$, $h_2$, and $h_3$, ordered by mass as $m_{h_1}<m_{h_2}<m_{h_3}$.
The lightest state $h_1$ is identified with the observed $125~\mathrm{GeV}$ Higgs boson, while $h_2$ is assumed to have a mass near $152~\mathrm{GeV}$.
Within this setup, we investigate the cascade $gg\to h_3\to h_2h_2\to(\gamma\gamma)(\chi\chi)$ as an explanation of the $\gamma\gamma+\mathrm{MET}$ excesses, while requiring $\chi$ to reproduce the full observed DM relic density through thermal freeze-out.
To allow both steps of the cascade to proceed on shell, the analysis is restricted to $m_{h_3}>2m_{h_2}$ and $m_\chi<m_{h_2}/2$.
At fixed $m_{h_2}$, increasing $m_{h_3}$ gives the daughter scalars larger momenta in the $h_3$ rest frame and generally hardens the MET spectrum, thereby affecting the signal acceptance in the Low-MET and High-MET categories.

The same scalar sector also provides a connection to gravitational waves (GWs).
Interactions between the doublets and the singlet can modify the finite-temperature potential and support first-order cosmological phase transitions, and the resulting bubble dynamics can generate a stochastic GW background~\cite{Zhang:2021alu,Caprini:2019egz}.
Motivated by this possibility, we explore whether the parameter regions accommodating the $\gamma\gamma+\mathrm{MET}$ excesses near $152~\mathrm{GeV}$ and the full observed DM relic density can also support first-order phase transitions with potentially detectable GWs.
The thermal history, bubble nucleation, percolation, and transition completion are investigated, while the GW spectra from sound waves and turbulence are evaluated, taking into account the finite lifetime of the acoustic source~\cite{Ellis:2020awk,Guo:2020grp}.
Our study thus examines a common scalar-sector origin for the proposed $\gamma\gamma+\mathrm{MET}$ signal near $152~\mathrm{GeV}$, the cosmological DM relic density, and potentially observable phase-transition GWs.

The paper is organized as follows.
In Sec.~\ref{sec:model}, we introduce the S2HDM and present its scalar spectrum and relevant couplings.
Section~\ref{sec:scan} describes the parameter scan and the theoretical and experimental constraints, followed by an analysis of the $\gamma\gamma+\mathrm{MET}$ excesses and DM phenomenology.
In Sec.~\ref{sec:pt_gw}, we investigate the thermal phase-transition history and the resulting GW signals, including their detectability at future space-based interferometers.
Finally, we draw our conclusions in Sec.~\ref{sec:conclusions}.

\section{The singlet-extended two-Higgs-doublet model}
\label{sec:model}
The 2HDM extends the SM by introducing an additional $SU(2)_L$ scalar doublet.
Under the assumption of natural flavor conservation, four standard types of Yukawa assignments can be constructed to avoid tree-level flavor-changing neutral currents~\cite{Glashow:1976nt,Paschos:1976ay,Branco:2011iw}.
In this work, we consider the Type-I 2HDM, in which all SM fermions couple to the same Higgs doublet, denoted by $\Phi_2$.
We further extend the model by introducing a complex scalar singlet $S$, providing a framework for pNGB DM~\cite{Jiang:2019soj,Biekotter:2021ovi,Zhang:2021alu,Biekotter:2022bxp}.
The two Higgs doublets carry hypercharge $Y=1/2$, whereas $S$ has $Y=0$.

The singlet sector is assumed to possess a global $U(1)$ symmetry, $S\to e^{i\alpha}S$, which is softly broken to a residual $Z_2$ symmetry, $S\to -S$, by a quadratic term.
In the doublet sector, a separate $Z_2$ symmetry, $\Phi_1\to-\Phi_1$ with $\Phi_2$ unchanged, is imposed on the scalar potential and softly broken only by the quadratic mixing term proportional to $m_{12}^2$~\cite{Ginzburg:2004vp}.
Under these assumptions, the full CP-conserving scalar potential consists of a $U(1)$-symmetric part $V_{\Phi_i,S}$ and a soft-breaking part $V_{\mathrm{soft}}$, with $V=V_{\Phi_i,S}+V_{\mathrm{soft}}$.
The $U(1)$-symmetric part is given by~\cite{Jiang:2019soj,Zhang:2021alu}
\begin{equation}
\begin{aligned}
V_{\Phi_i,S}={}&
m_{11}^2|\Phi_1|^2+m_{22}^2|\Phi_2|^2
-m_{12}^2\left(\Phi_1^\dagger\Phi_2+\mathrm{H.c.}\right)
+\frac{\lambda_1}{2}|\Phi_1|^4+\frac{\lambda_2}{2}|\Phi_2|^4
\\
&+\lambda_3|\Phi_1|^2|\Phi_2|^2
+\lambda_4|\Phi_1^\dagger\Phi_2|^2
+\frac{\lambda_5}{2}\left[(\Phi_1^\dagger\Phi_2)^2+\mathrm{H.c.}\right]
\\
&-m_S^2|S|^2+\frac{\lambda_S}{2}|S|^4
+\kappa_1|\Phi_1|^2|S|^2+\kappa_2|\Phi_2|^2|S|^2 .
\end{aligned}
\label{eq:scalar_potential}
\end{equation}
Here, the first two lines constitute the general CP-conserving 2HDM potential with a softly broken $Z_2$ symmetry, while the third line contains the $U(1)$-symmetric singlet potential and its interactions with the two Higgs doublets.
The soft $U(1)$-breaking term is given by
\begin{equation}
V_{\mathrm{soft}}
=
-\frac{m_S^{\prime\,2}}{4}
\left(S^2+S^{*2}\right).
\label{eq:soft_potential}
\end{equation}

The model contains two complex scalar $SU(2)_L$ doublets, $\Phi_1$ and $\Phi_2$, and a complex scalar singlet $S$, which are parametrized as
\begin{equation}
\Phi_1=
\begin{pmatrix}
\phi_1^+\\[2mm]
\dfrac{v_1+\rho_1+i\eta_1}{\sqrt{2}}
\end{pmatrix},
\qquad
\Phi_2=
\begin{pmatrix}
\phi_2^+\\[2mm]
\dfrac{v_2+\rho_2+i\eta_2}{\sqrt{2}}
\end{pmatrix},
\qquad
S=\frac{1}{\sqrt{2}}\left(v_s+s+i\chi\right).
\label{eq:field_expansions}
\end{equation}
Here, $v_1$, $v_2$, and $v_s$ denote the real VEVs of the two Higgs doublets and the singlet, respectively.
The electroweak VEV and the ratio of the doublet VEVs are defined as
\begin{equation}
v=\sqrt{v_1^2+v_2^2}\simeq246~\mathrm{GeV},
\qquad
\tan\beta=\frac{v_2}{v_1},
\label{eq:vev_definitions}
\end{equation}
such that $v_1=v\cos\beta$ and $v_2=v\sin\beta$.
The tree-level stationarity conditions at the vacuum allow the quadratic parameters $m_{11}^2$, $m_{22}^2$, and $m_S^2$ to be expressed as
\begin{equation}
\begin{aligned}
m_{11}^2 &= m_{12}^2\tan\beta-\frac{1}{2}\left(\lambda_1v_1^2+\lambda_{345}v_2^2+\kappa_1v_s^2\right),\\
m_{22}^2 &= m_{12}^2\cot\beta-\frac{1}{2}\left(\lambda_2v_2^2+\lambda_{345}v_1^2+\kappa_2v_s^2\right),\\
m_S^2 &= \frac{1}{2}\left(\kappa_1v_1^2+\kappa_2v_2^2+\lambda_Sv_s^2-m_S^{\prime\,2}\right),
\end{aligned}
\label{eq:stationarity_conditions}
\end{equation}
where $\lambda_{345}\equiv\lambda_3+\lambda_4+\lambda_5$.
The singlet vacuum preserves the conjugation symmetry $S\to S^*$, under which $\chi\to-\chi$, ensuring the stability of the pNGB DM candidate $\chi$.
Its tree-level mass is given by
\begin{equation}
m_\chi^2=m_S^{\prime\,2}.
\label{eq:chi_mass}
\end{equation}

After imposing these conditions, the tree-level mass-squared matrix in the CP-even basis $(\rho_1,\rho_2,s)$ takes the form
\begin{equation}
\mathcal{M}_{\mathrm{even}}^2=
\begin{pmatrix}
\lambda_1v_1^2+m_{12}^2\tan\beta &
\lambda_{345}v_1v_2-m_{12}^2 &
\kappa_1v_1v_s\\
\lambda_{345}v_1v_2-m_{12}^2 &
\lambda_2v_2^2+m_{12}^2\cot\beta &
\kappa_2v_2v_s\\
\kappa_1v_1v_s &
\kappa_2v_2v_s &
\lambda_Sv_s^2
\end{pmatrix}.
\label{eq:cp_even_mass_matrix}
\end{equation}
The three CP-even mass eigenstates are related to the interaction eigenstates by a real orthogonal matrix $R$,
\begin{equation}
\begin{pmatrix}
h_1\\h_2\\h_3
\end{pmatrix}
=
R
\begin{pmatrix}
\rho_1\\\rho_2\\s
\end{pmatrix},
\qquad
R\mathcal{M}_{\mathrm{even}}^2R^T
=
\operatorname{diag}\left(m_{h_1}^2,m_{h_2}^2,m_{h_3}^2\right).
\label{eq:cp_even_diagonalization}
\end{equation}
The three CP-even mass eigenstates are ordered by increasing mass, $m_{h_1}<m_{h_2}<m_{h_3}$, and the mixing matrix is parametrized by three angles $\alpha_1$, $\alpha_2$, and $\alpha_3$ as
\begin{equation}
R=
\begin{pmatrix}
c_1c_2 & s_1c_2 & s_2\\
-c_1s_2s_3-s_1c_3 & c_1c_3-s_1s_2s_3 & c_2s_3\\
-c_1s_2c_3+s_1s_3 & -c_1s_3-s_1s_2c_3 & c_2c_3
\end{pmatrix},
\label{eq:mixing_matrix}
\end{equation}
where $c_i\equiv\cos\alpha_i$ and $s_i\equiv\sin\alpha_i$.

The CP-odd and charged mass eigenstates arising from the doublet sector are given by
\begin{equation}
A=-\sin\beta\,\eta_1+\cos\beta\,\eta_2,
\qquad
H^\pm=-\sin\beta\,\phi_1^\pm+\cos\beta\,\phi_2^\pm,
\label{eq:odd_charged_states}
\end{equation}
while the orthogonal combinations are the Goldstone modes associated with the longitudinal components of the $Z$ and $W^\pm$ bosons, respectively.
The tree-level squared masses of $A$ and $H^\pm$ are given by
\begin{equation}
\begin{aligned}
m_A^2 &= \frac{m_{12}^2}{\sin\beta\cos\beta}-\lambda_5v^2,\\
m_{H^\pm}^2 &= \frac{m_{12}^2}{\sin\beta\cos\beta}-\frac{\lambda_4+\lambda_5}{2}v^2.
\end{aligned}
\label{eq:odd_charged_masses}
\end{equation}

The scalar sector of the model can be parametrized by the following twelve independent input parameters:
\begin{equation}
\left\{
m_{h_1},m_{h_2},m_{h_3},m_A,m_{H^\pm},m_\chi,
\alpha_1,\alpha_2,\alpha_3,\tan\beta,v_s,m_{12}^2
\right\}.
\label{eq:physical_parameters}
\end{equation}
The quartic couplings in the scalar potential can be reconstructed from these inputs using Eqs.~\eqref{eq:cp_even_mass_matrix}, \eqref{eq:cp_even_diagonalization}, and \eqref{eq:odd_charged_masses}.
In particular, the doublet--singlet portal couplings $\kappa_1$ and $\kappa_2$ can be expressed in terms of the physical masses and mixing matrix elements as
\begin{equation}
\begin{aligned}
\kappa_1 &= \frac{1}{v v_s\cos\beta}\sum_{i=1}^{3}m_{h_i}^2R_{i1}R_{i3},\\
\kappa_2 &= \frac{1}{v v_s\sin\beta}\sum_{i=1}^{3}m_{h_i}^2R_{i2}R_{i3}.
\end{aligned}
\label{eq:portal_couplings}
\end{equation}

The mixing matrix also determines the reduced couplings of the CP-even Higgs bosons to SM particles.
In the Type-I Yukawa configuration, these couplings are given by
\begin{equation}
C_{h_iVV}\equiv\frac{g_{h_iVV}}{g_{h_{\mathrm{SM}}VV}}
=\cos\beta\,R_{i1}+\sin\beta\,R_{i2},
\qquad
C_{h_i ff}\equiv\frac{g_{h_i ff}}{g_{h_{\mathrm{SM}}ff}}
=\frac{R_{i2}}{\sin\beta},
\label{eq:normalized_higgs_couplings}
\end{equation}
where $i=1,2,3$, $V=W,Z$, and $f$ denotes any charged SM fermion.
The singlet component has no direct tree-level coupling to these SM fields and contributes through scalar mixing.

The interactions between the CP-even scalars and the DM particle are defined by
\begin{equation}
\mathcal{L}\supset-\frac{1}{2}\sum_{i=1}^{3}g_{h_i\chi\chi}\,h_i\chi^2.
\label{eq:dm_interaction}
\end{equation}
At tree level, the corresponding couplings satisfy
\begin{equation}
g_{h_i\chi\chi}
=\kappa_1v_1R_{i1}+\kappa_2v_2R_{i2}+\lambda_Sv_sR_{i3}
=\frac{m_{h_i}^2}{v_s}R_{i3},
\label{eq:dm_higgs_coupling}
\end{equation}
where the last equality follows from the diagonalization of the CP-even mass-squared matrix.
These interactions mediate DM annihilation and allow the invisible decays $h_i\to\chi\chi$ whenever $m_{h_i}>2m_\chi$.

The diphoton decay of a CP-even Higgs boson is induced at one loop by charged fermions, $W$ bosons, and charged Higgs bosons.
The corresponding partial decay width is given by~\cite{Posch:2010hx,Djouadi:2005gj}
\begin{equation}
\Gamma(h_i\to\gamma\gamma)
=
\frac{\alpha_{\mathrm{em}}^2m_{h_i}^3}{256\pi^3v^2}
\left|
\sum_f N_c^f Q_f^2 C_{h_i ff}A_{1/2}(\tau_f)
+C_{h_i VV}A_1(\tau_W)
+\frac{v g_{h_iH^+H^-}}{2m_{H^\pm}^2}A_0(\tau_{H^\pm})
\right|^2,
\label{eq:diphoton_width}
\end{equation}
where $\tau_x=m_{h_i}^2/(4m_x^2)$, and $N_c^f$ and $Q_f$ denote the color factor and electric charge of the fermion $f$, respectively.
The trilinear coupling $g_{h_iH^+H^-}$ is defined through
\begin{equation}
\mathcal{L}\supset-\sum_{i=1}^{3}g_{h_iH^+H^-}h_iH^+H^-.
\label{eq:charged_higgs_trilinear}
\end{equation}
The functions $A_{1/2}$, $A_1$, and $A_0$ describe the spin-$1/2$, spin-$1$, and spin-$0$ loop contributions, respectively, and their explicit expressions can be found in Ref.~\cite{Branco:2011iw}.
The charged-Higgs contribution can interfere constructively or destructively with the fermion and $W$-boson contributions, thereby modifying the diphoton decay width.

\section{Collider and dark matter phenomenology}
\label{sec:scan}

The model is implemented in \texttt{SARAH-4.15.3}~\cite{Staub:2013tta}, which generates the model-specific routines for \texttt{SPheno-4.0.6}~\cite{Porod:2003um,Porod:2011nf}.
The resulting code is used to calculate the mass spectrum and decay observables throughout the parameter scan.
We identify $h_1$ with the observed SM-like Higgs boson and fix $m_{h_1}=125.09~\mathrm{GeV}$, while focusing on scenarios with $m_{h_2}$ near $152~\mathrm{GeV}$.
The main scan uses the following parameter ranges:
\begin{equation}
\begin{aligned}
&m_{h_2}\in[147,157]~\mathrm{GeV},
&&m_{h_3}\in[360,500]~\mathrm{GeV},\\
&m_A\in[400,550]~\mathrm{GeV},
&&m_\chi\in[65,78]~\mathrm{GeV},\\
&\Delta_m\in[-10,20]~\mathrm{GeV},
&&M\in[80,200]~\mathrm{GeV},\\
&v_s\in[1000,2500]~\mathrm{GeV},
&&\tan\beta\in[1.5,3],\\
&\sin(\alpha_1-\beta)\in[0,0.2],
&&\alpha_2\in[0,\pi/9],\\
&\alpha_3\in[\pi/12,\pi/3].
\end{aligned}
\label{eq:scan_ranges}
\end{equation}
Here, $\Delta_m=m_{H^\pm}-m_A$ and $M^2=m_{12}^2/(\sin\beta\cos\beta)$.
All mixing angles are expressed in radians.
The final dataset also includes retained samples from supplementary random scans and neighborhood MCMC searches extending beyond these initial ranges.
We additionally require $2m_\chi<m_{h_2}$ to ensure that the invisible decay $h_2\to\chi\chi$ is kinematically allowed.

To identify phenomenologically viable parameter regions, we impose the following theoretical and experimental constraints.

\begin{enumerate}

\item \textit{Perturbativity and perturbative unitarity.}
We impose perturbativity bounds on the quartic couplings,
\begin{equation}
|\lambda_i|<4\pi\quad(i=1,\ldots,5),
\qquad
|\lambda_S|<4\pi,
\qquad
|\kappa_{1,2}|<4\pi.
\label{eq:perturbativity}
\end{equation}
Tree-level perturbative unitarity is tested using the scalar-scattering routines generated by \texttt{SARAH-4.15.3}~\cite{Staub:2013tta} and implemented in \texttt{SPheno-4.0.6}~\cite{Goodsell:2018tti}.
The eigenvalues of the zeroth partial-wave scattering matrix are required to satisfy
\begin{equation}
\left|\operatorname{Re}a_{0,k}\right|<\frac{1}{2}.
\label{eq:perturbative_unitarity}
\end{equation}
We require both the high-energy and finite-energy unitarity tests to be satisfied.
The latter includes contributions from trilinear scalar interactions and uses five energy-scan steps between $1$ and $2~\mathrm{TeV}$.

\item \textit{Boundedness from below.}
We require the scalar potential to be bounded from below by imposing the copositivity conditions on its quartic part~\cite{Kannike:2012pe}.
These conditions account for both single-field and mixed-field directions involving the two Higgs doublets and the complex singlet.

\item \textit{Electroweak precision constraints.}
The oblique parameters $S$, $T$, and $U$ are calculated with \texttt{SPheno}.
We adopt the three-parameter fit from PDG 2026~\cite{ParticleDataGroup:2026mpi},
\begin{equation}
S=0.018\pm0.095,\qquad T=0.040\pm0.120,\qquad U=-0.013\pm0.086,
\end{equation}
with correlation coefficients $\rho_{ST}=0.91$, $\rho_{SU}=-0.66$, and $\rho_{TU}=-0.88$.
We require $\chi^2_{STU}<7.815$, corresponding to the joint $95\%$ confidence region for three degrees of freedom.

\item \textit{Vacuum stability.}
The vacuum-stability constraint is imposed on all samples retained to explain at least one of the two excesses within its individual $1\sigma$ interval and reproduce the observed DM relic density within the assumed theoretical uncertainty.
We examine the zero-temperature vacuum structure at tree level using \texttt{Vevacious++}, the C++ implementation of the \texttt{Vevacious} framework~\cite{Camargo-Molina:2013qva}, including neutral, charge-breaking, CP-breaking, and dark-field directions.
The electroweak vacuum is required to be either the global minimum of the scalar potential or a metastable minimum with a lifetime longer than the age of the Universe.

\item \textit{Flavor constraints.}
Flavor observables are evaluated using \texttt{FlavorKit}~\cite{Porod:2014xia} within the \texttt{SARAH}/\texttt{SPheno} framework.
For $B\to X_s\gamma$ with $E_\gamma>1.6~\mathrm{GeV}$, the SM contribution is updated using the central value $3.54\times10^{-4}$ from Ref.~\cite{Misiak:2026sqy}, while the new-physics contribution calculated with \texttt{FlavorKit} is retained.
We impose the following experimental $2\sigma$ intervals based on the PDG 2026 averages~\cite{ParticleDataGroup:2026mpi}:
\begin{equation}
\begin{array}{rcl}
3.11\times10^{-4} & \leq &
\mathrm{BR}(B\to X_s\gamma)\leq 3.87\times10^{-4},\\[2mm]
2.80\times10^{-9} & \leq &
\mathrm{BR}(B_s\to\mu^+\mu^-)\leq 3.88\times10^{-9}.
\end{array}
\label{eq:flavor_constraints}
\end{equation}

\item \textit{Higgs-sector constraints.}
Searches for additional Higgs bosons and measurements of the observed Higgs boson are tested using the \texttt{HiggsBounds}~\cite{Bechtle:2020pkv} and \texttt{HiggsSignals}~\cite{Bechtle:2020uwn} components of \texttt{HiggsTools-1.1.3}~\cite{Bahl:2022igd}.
Samples excluded by \texttt{HiggsBounds} at the $95\%$ confidence level are discarded.
Compatibility with the measured properties of the observed Higgs boson is assessed using \texttt{HiggsSignals}, with the requirement
\begin{equation}
\chi^2_{\mathrm{HS}}\leq 189.42.
\end{equation}
This threshold corresponds to a $95\%$ goodness-of-fit criterion, assuming $\nu=N_{\mathrm{obs}}=159$ degrees of freedom.

\item \textit{Dark matter constraints.}
The relic density and DM--nucleon scattering cross sections are calculated using \texttt{micrOMEGAs-6.2.4}~\cite{Alguero:2023zol}.
Adopting the Planck 2015 reference value $\Omega_{\mathrm{DM}}h^2=0.1188$~\cite{Planck:2015fie} and allowing for a $10\%$ theoretical uncertainty in the relic-density calculation, we impose the upper bound $\Omega_\chi h^2\leq0.13068$.
No lower bound is imposed, allowing $\chi$ to constitute only a fraction of the total DM.
For the subset in which $\chi$ accounts for the full observed DM relic density, we require $0.10692\leq\Omega_\chi h^2\leq0.13068$.
The tree-level spin-independent scattering predictions are compared with the observed $90\%$ confidence-level limits reported by the LUX-ZEPLIN (LZ) experiment in 2025~\cite{LZ:2024zvo}.
For this comparison, we use the larger of the proton and neutron cross sections without relic-density rescaling.

\end{enumerate}

\begin{figure*}[!ht]
    \centering
    \includegraphics[width=0.3\linewidth]{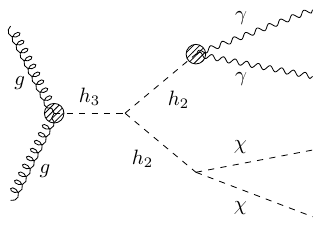}
    \caption{
Representative Feynman diagram for the signal process $gg\to h_3\to h_2h_2\to(\gamma\gamma)(\chi\chi)$, yielding a $\gamma\gamma+\mathrm{MET}$ signature.
The shaded circles represent the loop-induced $ggh_3$ and $h_2\gamma\gamma$ vertices.
}
\label{fig:feny}

\end{figure*}

To interpret the $\gamma\gamma+\mathrm{MET}$ excesses near $152~\mathrm{GeV}$, we consider the cascade process $gg\to h_3\to h_2h_2\to(\gamma\gamma)(\chi\chi)$, as illustrated in Fig.~\ref{fig:feny}.
The gluon-fusion production cross section of $h_3$ at $\sqrt{s}=13~\mathrm{TeV}$ is obtained using \texttt{HiggsTools-1.1.3}~\cite{Bahl:2022igd}.
In the effective-coupling approximation, it can be expressed as
\begin{equation}
\sigma(gg\to h_3)
\simeq
|C_{h_3gg}|^2\,\sigma_{\mathrm{SM}}^{gg\mathrm{F}}(m_{h_3}),
\label{eq:h3_production}
\end{equation}
where $C_{h_3gg}$ denotes the effective coupling of $h_3$ to gluons normalized to that of an SM Higgs boson of the same mass.
The reference cross section $\sigma_{\mathrm{SM}}^{gg\mathrm{F}}(m_{h_3})$ is taken from the NNLO+NNLL gluon-fusion predictions in the CERN YR4 tables~\cite{LHCHiggsCrossSectionWorkingGroup:2016ypw}.
In the narrow-width approximation, the signal cross section before event selection is then given by
\begin{equation}
\begin{aligned}
\sigma_s={}&
2\,\sigma(gg\to h_3)\,
\mathrm{BR}(h_3\to h_2h_2)\mathrm{BR}(h_2\to\gamma\gamma)\,
\mathrm{BR}(h_2\to\chi\chi).
\end{aligned}
\label{eq:signal_cross_section}
\end{equation}
The factor of two accounts for the two possible assignments of the diphoton and invisible decays to the two $h_2$ bosons.

\begin{figure*}[!ht]
    \centering
    \setlength{\abovecaptionskip}{-5pt}
    \includegraphics[width=\linewidth]{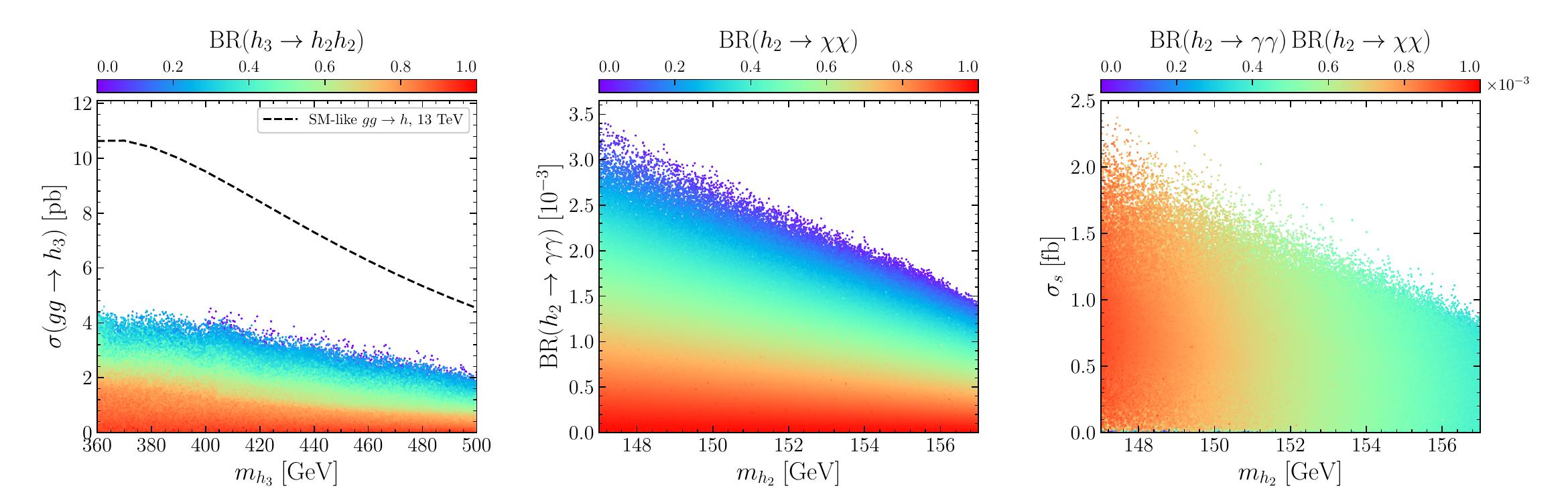}
    \caption{
Surviving samples in the $m_{h_3}$--$\sigma(gg\to h_3)$ plane (left), the $m_{h_2}$--$\mathrm{BR}(h_2\to\gamma\gamma)$ plane (middle), and the $m_{h_2}$--$\sigma_s$ plane (right).
The colors indicate $\mathrm{BR}(h_3\to h_2h_2)$, $\mathrm{BR}(h_2\to\chi\chi)$, and $\mathrm{BR}(h_2\to\gamma\gamma)\mathrm{BR}(h_2\to\chi\chi)$ in the left, middle, and right panels, respectively.
The black dashed curve in the left panel denotes the gluon-fusion production cross section for an SM-like Higgs boson of the same mass at $\sqrt{s}=13~\mathrm{TeV}$.
}
\label{fig:figure1}
\end{figure*}

The random-scan component of the displayed dataset contains approximately $8.8\times10^5$ samples after the initial selection.
We further explore the parameter space using Markov chain Monte Carlo (MCMC) searches around phenomenologically relevant samples and samples with previously obtained GW signals.
Combining the random-scan, MCMC, and additional archived samples yields approximately $1.3\times10^6$ samples, shown in Fig.~\ref{fig:figure1}.
The additional vacuum-stability requirement is applied to the subset used in the phase-transition and GW analysis.
From left to right, the three panels display the samples in the $m_{h_3}$--$\sigma(gg\to h_3)$, $m_{h_2}$--$\mathrm{BR}(h_2\to\gamma\gamma)$, and $m_{h_2}$--$\sigma_s$ planes.
The colors indicate $\mathrm{BR}(h_3\to h_2h_2)$, $\mathrm{BR}(h_2\to\chi\chi)$, and $\mathrm{BR}(h_2\to\gamma\gamma)\mathrm{BR}(h_2\to\chi\chi)$, respectively.

As shown in the left panel of Fig.~\ref{fig:figure1}, the upper envelope of $\sigma(gg\to h_3)$ decreases for $m_{h_3}\gtrsim400~\mathrm{GeV}$, falling from approximately $4.5~\mathrm{pb}$ at $m_{h_3}=400~\mathrm{GeV}$ to $2.2~\mathrm{pb}$ at $m_{h_3}=500~\mathrm{GeV}$.
In addition, larger production cross sections are generally associated with smaller values of $\mathrm{BR}(h_3\to h_2h_2)$, and this anticorrelation limits the attainable production rate of $gg\to h_3\to h_2h_2$.
The middle panel shows that the upper envelope of $\mathrm{BR}(h_2\to\gamma\gamma)$ decreases from approximately $0.35\%$ to $0.15\%$ as $m_{h_2}$ increases from $147$ to $157~\mathrm{GeV}$.
In contrast, the invisible branching ratio $\mathrm{BR}(h_2\to\chi\chi)$ can approach unity throughout the considered mass range.
The right panel shows that the maximum attainable signal cross section $\sigma_s$ also decreases with increasing $m_{h_2}$.
This trend is associated with the larger branching-ratio products $\mathrm{BR}(h_2\to\gamma\gamma)\mathrm{BR}(h_2\to\chi\chi)$ attainable at lower $m_{h_2}$, which favor larger signal rates.

In our scenario, explaining the $\gamma\gamma+\mathrm{MET}$ excesses requires not only an $h_2$ mass near $152~\mathrm{GeV}$ but also suitable signal kinematics, which depend strongly on the mass of the parent scalar $h_3$.
For a fixed $m_{h_2}$, increasing $m_{h_3}$ gives the daughter scalars larger momenta in the $h_3$ rest frame, generally leading to a harder MET spectrum.
To quantify this dependence and determine the signal acceptance and selection efficiencies, we perform a collider simulation of the signal process.

For the collider simulation, we generate parton-level signal events in proton--proton collisions at $\sqrt{s}=13~\mathrm{TeV}$ using \texttt{MadGraph5\_aMC@NLO-3.8.0}~\cite{Alwall:2014hca} at leading order, with the scalar cascade implemented in the narrow-width approximation through effective $ggh_3$ and $h_2\gamma\gamma$ vertices.
The generated events are passed to \texttt{Pythia-8.318}~\cite{Bierlich:2022pfr} for parton showering and hadronization.
The detector response is then simulated using \texttt{Delphes-3.5.1}~\cite{deFavereau:2013fsa} with the ATLAS card \texttt{delphes\_card\_ATLAS.tcl}, treating $\chi$ as a stable invisible particle.
Jets are reconstructed with the anti-$k_t$ algorithm~\cite{Cacciari:2008gp} implemented in \texttt{FastJet}~\cite{Cacciari:2011ma}, using a radius parameter of $R=0.4$.
To investigate the dependence of the signal kinematics on the heavy-scalar mass, we fix $m_{h_2}=152~\mathrm{GeV}$ and $m_\chi=71~\mathrm{GeV}$ while varying $m_{h_3}$ from $360$ to $500~\mathrm{GeV}$ in steps of $20~\mathrm{GeV}$.
At each mass value, $5\times10^4$ signal events are simulated.
These events are used in the subsequent analysis to calculate the selection efficiencies and assess whether the model can account for the observed excesses.

\begin{figure*}[!ht]
    \centering
    \setlength{\abovecaptionskip}{-5pt}
    \includegraphics[width=\linewidth]{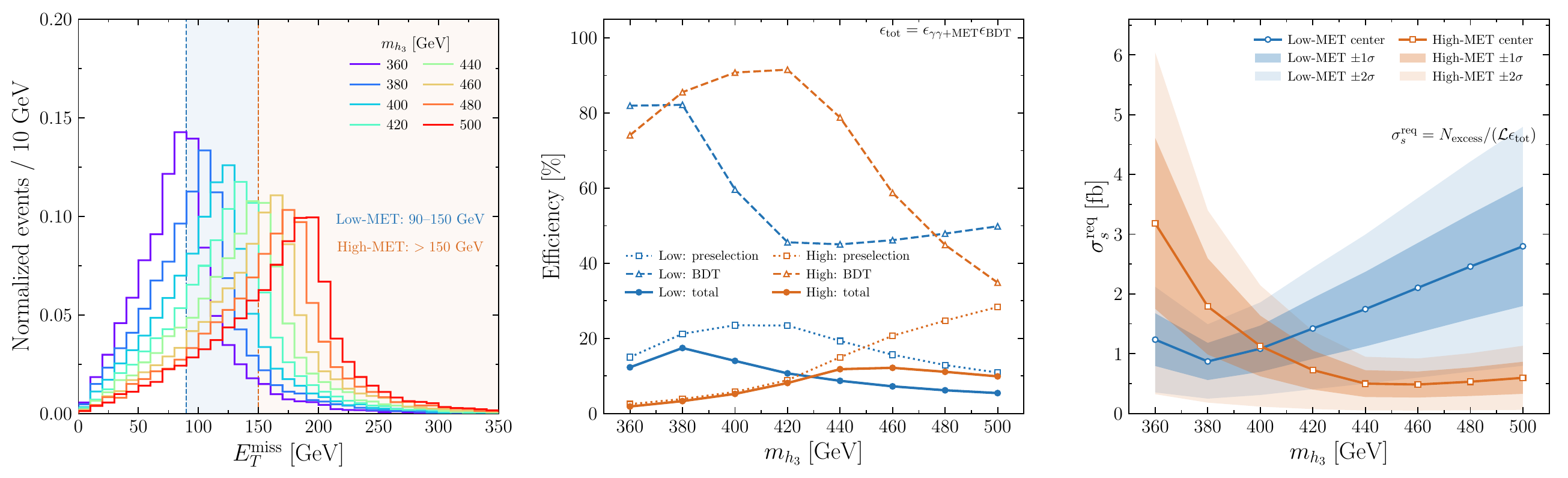}
\caption{Normalized missing transverse momentum distributions (left), selection efficiencies (middle), and required signal cross sections (right) for $gg\to h_3\to h_2h_2\to\gamma\gamma\chi\chi$, with $m_{h_2}=152~\mathrm{GeV}$ and $m_\chi=71~\mathrm{GeV}$.
In the left panel, the curves correspond to different $h_3$ masses, while the blue and orange shaded regions denote the Low-MET ($90<E_T^{\mathrm{miss}}<150~\mathrm{GeV}$) and High-MET ($E_T^{\mathrm{miss}}>150~\mathrm{GeV}$) regions, respectively.
In the middle panel, dotted, dashed, and solid lines represent the preselection, conditional BDT-loose, and total efficiencies, respectively, with blue and orange corresponding to the Low-MET and High-MET categories.
In the right panel, the blue and orange curves show the signal cross sections required to reproduce the central excess yields in the corresponding BDT-loose categories reported in Ref.~\cite{Bhattacharya:2025rfr}, using $\mathcal{L}=139~\mathrm{fb}^{-1}$.
The darker and lighter shaded bands indicate the $1\sigma$ and $2\sigma$ intervals obtained by propagating the quoted uncertainties on these yields.}
    \label{fig:figure2}
\end{figure*}

The left panel of Fig.~\ref{fig:figure2} shows the normalized MET distributions for different values of $m_{h_3}$ after the photon selection and lepton veto, but before imposing the regional MET or boosted decision tree (BDT) requirements.
The photon selection follows the fiducial acceptance, transverse-energy requirements, and diphoton invariant-mass window of Ref.~\cite{ATLAS:2021jbf}.
Object reconstruction and identification, photon isolation, the lepton veto, and object overlap removal are treated within our Delphes-based fast-simulation approximation.
Pile-up interactions and primary-vertex misassignment are not simulated.
Each MET distribution is normalized to the total number of events passing the implemented photon selection and lepton veto.
The blue and orange shaded bands indicate the Low-MET region, $90<E_T^{\mathrm{miss}}<150~\mathrm{GeV}$, and the High-MET region, $E_T^{\mathrm{miss}}>150~\mathrm{GeV}$, respectively.
As expected, increasing $m_{h_3}$ from $360$ to $500~\mathrm{GeV}$ hardens the MET distribution, shifting its peak from approximately $85$ to $195~\mathrm{GeV}$.
Consequently, lighter $h_3$ states preferentially populate the Low-MET region relative to the High-MET region, whereas heavier states enhance the relative acceptance of the High-MET selection.

The dotted curves in the middle panel of Fig.~\ref{fig:figure2} show the combined efficiencies of the implemented preselection requirements, including the photon selection, lepton veto, and regional MET cut, before applying the BDT selection.
For each region $r=\mathrm{Low},\mathrm{High}$, the preselection efficiency is defined as
\begin{equation}
\epsilon_{\gamma\gamma+\mathrm{MET}}^r=\frac{N_{\mathrm{pre}}^r}{N_{\mathrm{gen}}},
\label{eq:preselection_efficiency}
\end{equation}
where $N_{\mathrm{gen}}$ is the total number of generated signal events at each mass value and $N_{\mathrm{pre}}^r$ is the number passing the preselection requirements in region $r$.
In the Low-MET region (blue), the preselection efficiency increases from approximately $15.0\%$ at $m_{h_3}=360~\mathrm{GeV}$ to about $23.5\%$ around $400$--$420~\mathrm{GeV}$, before decreasing to $10.9\%$ at $500~\mathrm{GeV}$.
In the High-MET region (orange), it increases monotonically from $2.5\%$ to $28.4\%$ over the same mass range.

The BDT selection efficiency is estimated using the BDT-score map and category definitions provided in Ref.~\cite{ATLAS:2021jbf}.
The BDT score depends only on the diphoton transverse momentum $p_T^{\gamma\gamma}$ and the MET significance, defined as
\begin{equation}
S_{E_T^{\mathrm{miss}}}=\frac{E_T^{\mathrm{miss}}}{\sqrt{\sum E_T}},
\label{eq:met_significance}
\end{equation}
where $\sum E_T$ is the scalar sum of the transverse momenta of the reconstructed physics objects, including the soft contribution used in the MET calculation.
For each preselected simulated event, the values of these two variables are used to identify the corresponding BDT-score interval from the published score map.
The BDT selection efficiency in each MET region is then obtained as the fraction of preselected events satisfying the score requirement for the relevant category.
Specifically, the Low-MET and High-MET BDT-loose categories correspond to
\begin{equation}
\begin{aligned}
0.386 < \mathrm{BDT} < 0.864,
&\qquad \text{Low-MET BDT-loose},\\
0.694 < \mathrm{BDT} < 0.950,
&\qquad \text{High-MET BDT-loose}.
\end{aligned}
\label{eq:bdt_score_ranges}
\end{equation}
We consider the Low-MET and High-MET BDT-loose categories used in Ref.~\cite{Bhattacharya:2025rfr}.
For each region $r=\mathrm{Low},\mathrm{High}$, $N_{\mathrm{loose}}^r$ denotes the number of preselected simulated events that also satisfy the corresponding BDT-loose selection.
The BDT selection efficiency is evaluated relative to the preselected sample as
\begin{equation}
\epsilon_{\mathrm{BDT}}^r=\frac{N_{\mathrm{loose}}^r}{N_{\mathrm{pre}}^r}.
\label{eq:selection_efficiencies}
\end{equation}
Combining this with the preselection efficiency defined above, the total selection efficiency is
\begin{equation}
\epsilon_{\mathrm{tot}}^r=\epsilon_{\gamma\gamma+\mathrm{MET}}^r\epsilon_{\mathrm{BDT}}^r=\frac{N_{\mathrm{loose}}^r}{N_{\mathrm{gen}}}.
\label{eq:total_selection_efficiency}
\end{equation}
Preselected events outside the domain of the published BDT-score map remain unclassified and are included in $N_{\mathrm{pre}}^r$ but not in $N_{\mathrm{loose}}^r$.

As shown in the middle panel of Fig.~\ref{fig:figure2}, the conditional BDT-loose efficiency in the Low-MET region reaches approximately $82.2\%$ at $m_{h_3}=380~\mathrm{GeV}$, decreases to around $45\%$ at $420$--$440~\mathrm{GeV}$, and then rises to approximately $49.8\%$ at $500~\mathrm{GeV}$.
In contrast, the High-MET BDT-loose efficiency increases to approximately $91.5\%$ at $m_{h_3}=420~\mathrm{GeV}$ before decreasing to approximately $34.8\%$ at $500~\mathrm{GeV}$.
This decrease reflects the migration of higher-scoring events into the BDT-tight categories.
Combining these conditional efficiencies with the preceding selections, the total efficiency reaches approximately $17.4\%$ for Low-MET at $m_{h_3}=380~\mathrm{GeV}$ and $12.2\%$ for High-MET at $m_{h_3}=460~\mathrm{GeV}$.
Thus, although increasing $m_{h_3}$ enhances the High-MET acceptance, the competing BDT-loose requirement prevents the corresponding total efficiency from increasing monotonically.

Using these total efficiencies, we translate the fitted excess yields into the signal cross sections required to reproduce the two excesses within our model:
\begin{equation}
\sigma_s^{\mathrm{req},\,r}(m_{h_3})
=
\frac{N_{\mathrm{excess}}^{\,r}}
{\mathcal{L}\,\epsilon_{\mathrm{tot}}^{\,r}(m_{h_3})},
\qquad r=\mathrm{Low},\mathrm{High}.
\label{eq:required_signal_cross_section}
\end{equation}
Here, $\sigma_s^{\mathrm{req},\,r}$ denotes the required cross section for the $\gamma\gamma\chi\chi$ signal before event selection, and $\mathcal{L}=139~\mathrm{fb}^{-1}$ is the integrated luminosity.
We adopt the fitted signal yields $N_{\mathrm{excess}}^{\mathrm{Low}}=21.10\pm7.54$ and $N_{\mathrm{excess}}^{\mathrm{High}}=8.19\pm3.68$ reported in Ref.~\cite{Bhattacharya:2025rfr}.
The resulting cross sections are displayed in the right panel of Fig.~\ref{fig:figure2}, where the blue and orange curves correspond to Low-MET and High-MET, respectively.
The darker and lighter bands represent the $1\sigma$ and $2\sigma$ intervals obtained by propagating only the quoted uncertainties on the fitted excess yields, with the selection efficiencies held fixed.
Since the required cross section is inversely proportional to the total efficiency, the Low-MET curve reaches its minimum near $m_{h_3}=380~\mathrm{GeV}$ and then increases, whereas the High-MET curve exhibits a broad minimum around $440$--$460~\mathrm{GeV}$ before increasing at higher masses.
At $m_{h_3}=400~\mathrm{GeV}$, the two independently inferred central values are approximately $1.08$ and $1.13~\mathrm{fb}$, respectively, indicating that both excesses are compatible with a common signal normalization at this mass.
For the scanned samples, the lower and upper boundaries of the $1\sigma$ intervals for the required signal cross sections are obtained by linear interpolation in $m_{h_3}$ between the simulated mass points, using the reference samples with $m_{h_2}=152~\mathrm{GeV}$ and $m_\chi=71~\mathrm{GeV}$.

\begin{figure*}[!ht]
    \centering
    \setlength{\abovecaptionskip}{-5pt}
    \includegraphics[width=\linewidth]{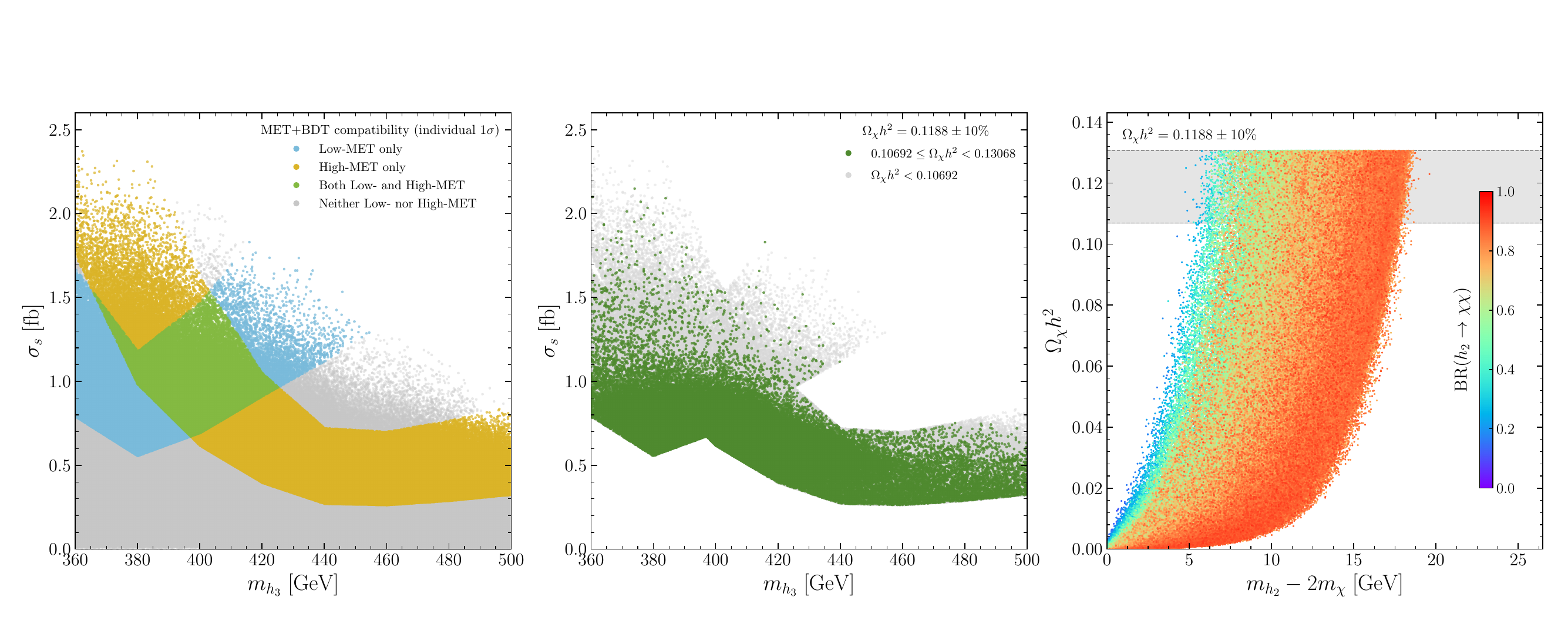}
    \caption{Surviving samples in the $\sigma_s$ versus $m_{h_3}$ plane (left and middle) and the $\Omega_\chi h^2$ versus $(m_{h_2}-2m_\chi)$ plane (right).
In the left panel, light-blue and yellow samples are compatible at $1\sigma$ with only the Low-MET and High-MET BDT-loose excesses, respectively, while light-green samples satisfy both criteria and gray samples satisfy neither.
The middle and right panels contain only the colored samples from the left panel.
In the middle panel, dark-green samples reproduce the observed DM relic density within the assumed $10\%$ theoretical uncertainty, whereas gray samples yield a smaller relic density.
In the right panel, the color represents $\mathrm{BR}(h_2\to\chi\chi)$, and the shaded horizontal band denotes the preferred relic-density interval.}
    \label{fig:figure3}
\end{figure*}

To assess the ability of the model to account for the excesses, we display the surviving samples in the $\sigma_s$ versus $m_{h_3}$ plane in the left panel of Fig.~\ref{fig:figure3}.
The light-blue and yellow samples are compatible at $1\sigma$ with only the Low-MET and High-MET BDT-loose excesses, respectively.
Samples compatible with both excesses are shown in light green, while those lying outside both $1\sigma$ intervals are shown in gray.
The results indicate that the model can accommodate either excess over an extended parameter region.
In particular, surviving samples populate the overlap of the two preferred regions, allowing both excesses to be explained simultaneously within their respective $1\sigma$ intervals.

\begin{table*}[!ht]
\centering
\caption{Input parameters and collider, DM, and flavor observables for five representative benchmark points.
We fix $m_{h_1}=125.09~\mathrm{GeV}$ and define $M^2=m_{12}^2/(\sin\beta\cos\beta)$.
Cross sections refer to $\sqrt{s}=13~\mathrm{TeV}$ before detector selection.}
\label{tab:benchmark_points}
\begingroup
\small
\setlength{\tabcolsep}{5pt}
\renewcommand{\arraystretch}{1.05}
\begin{tabular}{@{}lccccc@{}}
\toprule
 & Point 1 & Point 2 & Point 3 & Point 4 & Point 5 \\
\midrule
$\tan\beta$ & 2.5340 & 2.9084 & 2.6877 & 2.5251 & 2.7804 \\
$\sin(\alpha_1-\beta)$ & 0.1717 & 0.1412 & 0.1712 & 0.1789 & 0.1266 \\
$\alpha_2\ [\mathrm{rad}]$ & 0.2445 & 0.2452 & 0.2664 & 0.2308 & 0.2014 \\
$\alpha_3\ [\mathrm{rad}]$ & 0.4484 & 0.4931 & 0.4803 & 0.4290 & 0.3504 \\
$v_s\ [\mathrm{GeV}]$ & 1540.05 & 1681.93 & 1337.81 & 1559.32 & 1918.16 \\
$M\ [\mathrm{GeV}]$ & 169.62 & 186.00 & 170.51 & 187.88 & 178.03 \\
$m_{h_2}\ [\mathrm{GeV}]$ & 151.89 & 152.14 & 152.01 & 150.93 & 149.58 \\
$m_{h_3}\ [\mathrm{GeV}]$ & 398.51 & 396.42 & 397.48 & 389.64 & 411.93 \\
$m_A\ [\mathrm{GeV}]$ & 476.57 & 489.26 & 475.58 & 488.33 & 473.72 \\
$m_{H^\pm}\ [\mathrm{GeV}]$ & 476.00 & 484.64 & 477.33 & 487.58 & 484.91 \\
$m_\chi\ [\mathrm{GeV}]$ & 69.77 & 70.71 & 70.84 & 69.17 & 67.42 \\
$\sigma(gg\to h_3)\ [\mathrm{pb}]$ & 1.0051 & 0.9698 & 1.0826 & 0.9444 & 0.6654 \\
$\mathrm{BR}(h_3\to h_2h_2)$ & 0.5829 & 0.6051 & 0.6194 & 0.5158 & 0.5155 \\
$\mathrm{BR}(h_2\to\gamma\gamma)\ [10^{-3}]$ & 1.3086 & 1.2231 & 1.1242 & 1.4212 & 1.6555 \\
$\mathrm{BR}(h_2\to\chi\chi)$ & 0.4659 & 0.4845 & 0.5189 & 0.4715 & 0.4555 \\
$\sigma_s\ [\mathrm{fb}]$ & 0.7143 & 0.6956 & 0.7824 & 0.6528 & 0.5173 \\
$\Omega_\chi h^2$ & 0.11635 & 0.12023 & 0.12397 & 0.11435 & 0.12363 \\
$\mathrm{BR}(B\to X_s\gamma)\ [10^{-4}]$ & 3.4425 & 3.4672 & 3.4535 & 3.4442 & 3.4604 \\
$\mathrm{BR}(B_s\to\mu^+\mu^-)\ [10^{-9}]$ & 3.1451 & 3.1174 & 3.1327 & 3.1430 & 3.1249 \\
\bottomrule
\end{tabular}
\endgroup
\end{table*}

The samples compatible with at least one of the two excesses at $1\sigma$ are displayed again in the $\sigma_s$ versus $m_{h_3}$ plane in the middle panel of Fig.~\ref{fig:figure3}.
The dark-green samples also reproduce the observed DM relic density within the assumed $10\%$ theoretical uncertainty, whereas the gray samples yield a smaller relic density.
The right panel shows the same samples in the DM relic density $\Omega_\chi h^2$ versus $(m_{h_2}-2m_\chi)$ plane, with the color indicating $\mathrm{BR}(h_2\to\chi\chi)$ and the shaded band marking the preferred relic density interval.
Within the scanned parameter space, very small mass separations generally yield an underabundant relic density, while samples reproducing the observed relic density at larger separations tend to exhibit larger invisible branching ratios.
This behavior is qualitatively consistent with resonantly enhanced DM annihilation mediated by $h_2$ near $2m_\chi\simeq m_{h_2}$, which can substantially reduce the relic density~\cite{Griest:1990kh}.
Away from the resonance, maintaining efficient annihilation may require stronger interactions.
The invisible branching ratio also depends on phase space and competing decay modes.

Table~\ref{tab:benchmark_points} presents the input parameters and relevant observables for five representative benchmark points. 
Points 1, 2, and 3 simultaneously account for the $\gamma\gamma+\mathrm{MET}$ excesses in the Low-MET and High-MET regions within their respective $1\sigma$ intervals.
At the $1\sigma$ level, Point 4 accounts for only the Low-MET excess, whereas Point 5 accounts for only the High-MET excess.
All five points are consistent with the observed DM relic density within the assumed theoretical uncertainty.
The samples satisfying the relic-density requirement and explaining at least one of the two excesses within $1\sigma$ are selected for our subsequent analysis of cosmological phase transitions and the associated GW signals.

\section{Phase transitions and gravitational waves}
\label{sec:pt_gw}

Having identified parameter regions compatible with the $\gamma\gamma+\mathrm{MET}$ excesses and the observed DM relic density, we now investigate whether they can also give rise to first-order cosmological phase transitions and observable GW signals.
Such signals would provide a complementary probe of this scenario.

\subsection{Finite-temperature effective potential}
\label{subsec:finite_temperature_potential}

In the 2HDM, the connection between the scalar spectrum, vacuum structure, and thermal history has been extensively studied~\cite{Cline:1996mga,Basler:2016obg,Dorsch:2017nza}.
Here, we focus on phase transitions involving only the neutral CP-even background fields of the two Higgs doublets and the singlet.
The charged and CP-odd background fields are set to zero, and the remaining backgrounds are parametrized by three real fields $\boldsymbol{\varphi}=(\varphi_1,\varphi_2,\varphi_s)$ as
\begin{equation}
\Phi_i^{\mathrm{bg}}=\frac{1}{\sqrt{2}}\begin{pmatrix}0\\ \varphi_i\end{pmatrix},
\qquad
S^{\mathrm{bg}}=\frac{\varphi_s}{\sqrt{2}},
\qquad i=1,2.
\label{eq:thermal_backgrounds}
\end{equation}
At each temperature, the minima of the effective potential are determined by allowing the three background fields to vary independently.
The field values at a given minimum are denoted by $v_1(T)$, $v_2(T)$, and $v_s(T)$.

Including one-loop quantum and thermal corrections together with daisy resummation, the effective potential takes the form~\cite{Quiros:1999jp,Athron:2023xlk}
\begin{equation}
V_{\mathrm{eff}}(\boldsymbol{\varphi},T)
=
V_0(\boldsymbol{\varphi})
+V_{\mathrm{CW}}(\boldsymbol{\varphi})
+V_{\mathrm{CT}}(\boldsymbol{\varphi})
+V_T(\boldsymbol{\varphi},T)
+V_{\mathrm{ring}}(\boldsymbol{\varphi},T).
\label{eq:finite_temperature_potential}
\end{equation}
Here, $V_0$ is the tree-level potential obtained from Eqs.~\eqref{eq:scalar_potential} and \eqref{eq:soft_potential} by substituting the background fields in Eq.~\eqref{eq:thermal_backgrounds}.
The terms $V_{\mathrm{CW}}$ and $V_{\mathrm{CT}}$ denote the zero-temperature one-loop Coleman--Weinberg correction and the counterterm potential, respectively.
At finite temperature, $V_T$ describes the one-loop thermal correction, while $V_{\mathrm{ring}}$ accounts for the daisy-resummation contribution.

In the Landau gauge, the Coleman--Weinberg contribution is written as~\cite{Coleman:1973jx}
\begin{equation}
V_{\mathrm{CW}}(\boldsymbol{\varphi})
=
\frac{1}{64\pi^2}
\sum_i (-1)^{F_i}n_i\,m_i^4(\boldsymbol{\varphi})
\left[
\ln\frac{|m_i^2(\boldsymbol{\varphi})|}{Q^2}
-c_i
\right],
\label{eq:coleman_weinberg}
\end{equation}
where $m_i^2(\boldsymbol{\varphi})$ are the field-dependent squared masses, $n_i$ are positive multiplicity factors, and $F_i=0$ ($1$) for bosons (fermions).
The calculation includes contributions from all scalar degrees of freedom, including $\chi$ and the electroweak Goldstone modes, as well as from the electroweak gauge bosons and the top and bottom quarks.
The contributions from the remaining fermions are neglected because they have weak Yukawa couplings.
The renormalization scale is fixed at $Q=v$.
With transverse and longitudinal gauge polarizations counted separately, the constants are $c_i=1/2$ for transverse gauge modes and $c_i=3/2$ for longitudinal gauge modes, scalars, and fermions.

A polynomial counterterm potential is introduced to preserve the location of the zero-temperature electroweak vacuum and the CP-even curvature mass-squared matrix:
\begin{equation}
V_{\mathrm{CT}}(\boldsymbol{\varphi})
=
\sum_{a=1,2,s}\left(\delta\mu_a^2\varphi_a^2+\delta\lambda_a\varphi_a^4\right)
+\delta\lambda_{12}\varphi_1^2\varphi_2^2
+\delta\lambda_{1s}\varphi_1^2\varphi_s^2
+\delta\lambda_{2s}\varphi_2^2\varphi_s^2.
\label{eq:counterterm_potential}
\end{equation}
Its nine coefficients are fixed by the following renormalization conditions~\cite{Zhang:2021alu}:
\begin{equation}
\left.\frac{\partial(V_{\mathrm{CW}}+V_{\mathrm{CT}})}{\partial\varphi_a}\right|_{\boldsymbol{\varphi}=\boldsymbol{v}}=0,
\qquad
\left.\frac{\partial^2(V_{\mathrm{CW}}+V_{\mathrm{CT}})}{\partial\varphi_a\partial\varphi_b}\right|_{\boldsymbol{\varphi}=\boldsymbol{v}}=0,
\label{eq:counterterm_conditions}
\end{equation}
where $\boldsymbol{v}=(v_1,v_2,v_s)$ and $a,b\in\{1,2,s\}$.
For the electroweak Goldstone modes $G^0$ and $G^\pm$, the infrared-sensitive logarithms are regulated numerically by replacing $|m_i^2|$ with $\max(|m_i^2|,m_{h_1}^2)$, while leaving the $m_i^4$ prefactors unchanged.

The one-loop thermal contribution is given by~\cite{Dolan:1973qd}
\begin{equation}
V_T(\boldsymbol{\varphi},T)
=
\frac{T^4}{2\pi^2}
\left[
\sum_{i\in B}n_i
J_B\!\left(\frac{m_i^2(\boldsymbol{\varphi})}{T^2}\right)
-
\sum_{i\in F}n_i
J_F\!\left(\frac{m_i^2(\boldsymbol{\varphi})}{T^2}\right)
\right],
\label{eq:thermal_potential}
\end{equation}
where the bosonic and fermionic thermal functions are defined as
\begin{equation}
\begin{aligned}
J_B(y)&=\int_0^\infty dx\,x^2
\ln\left(1-e^{-\sqrt{x^2+y}}\right),\\
J_F(y)&=\int_0^\infty dx\,x^2
\ln\left(1+e^{-\sqrt{x^2+y}}\right).
\end{aligned}
\label{eq:thermal_functions}
\end{equation}

To account for the leading infrared-sensitive contributions from bosonic zero Matsubara modes~\cite{Carrington:1991hz}, we employ the Arnold--Espinosa prescription~\cite{Arnold:1992rz},
\begin{equation}
V_{\mathrm{ring}}(\boldsymbol{\varphi},T)
=
-\frac{T}{12\pi}
\sum_{i\in B_0}n_i\,
\operatorname{Re}
\left[
\left(\overline{m}_i^2(\boldsymbol{\varphi},T)\right)^{3/2}
-
\left(m_i^2(\boldsymbol{\varphi})\right)^{3/2}
\right].
\label{eq:ring_potential}
\end{equation}
Here, $B_0$ comprises the scalar modes and the longitudinal electroweak gauge modes, including the longitudinal photon.
The resummed squared masses $\overline{m}_i^2$ are obtained by diagonalizing the corresponding field-dependent mass matrices after adding their leading thermal self-energies,
\begin{equation}
\overline{m}_i^2(\boldsymbol{\varphi},T)
=
\operatorname{eig}_i
\left[
\mathcal{M}^2(\boldsymbol{\varphi})+\Pi(T)
\right].
\label{eq:thermal_masses}
\end{equation}
For the Type-I model, the expressions for the field-dependent mass matrices and thermal self-energies are taken from Ref.~\cite{Zhang:2021alu}.
Only the real part of the effective potential is retained when field-dependent squared masses become negative.
At the perturbative order considered, the predictions retain a residual dependence on the renormalization scale and the thermal-resummation prescription~\cite{Croon:2020cgk}.

\subsection{Phase-transition history and characteristic parameters}
\label{subsec:phase_transition_parameters}

To investigate the thermal history, we use \texttt{CosmoTransitions-2.0.2}~\cite{Wainwright:2011kj} to trace the temperature-dependent minima of the effective potential and calculate the multi-field tunneling action.
The three background fields are evolved independently, allowing thermal histories with a nonzero singlet VEV before electroweak symmetry breaking.
For a first-order transition, the critical temperature $T_c$ is determined by the degeneracy of the two competing minima,
\begin{equation}
V_{\mathrm{eff}}\!\left(\boldsymbol{\varphi}_{\mathrm{false}}(T_c),T_c\right)
=
V_{\mathrm{eff}}\!\left(\boldsymbol{\varphi}_{\mathrm{true}}(T_c),T_c\right).
\label{eq:critical_temperature}
\end{equation}
Here, $\boldsymbol{\varphi}_{\mathrm{false}}$ and $\boldsymbol{\varphi}_{\mathrm{true}}$ denote the false- and true-vacuum branches, respectively.
The realization of the transition is then assessed by calculating bubble nucleation and expansion.

Thermal tunneling is described by an $O(3)$-symmetric bounce solution $\boldsymbol{\varphi}(r)$, which represents the field configuration of a critical bubble.
The fields depend only on the radial distance $r$ from the bubble center.
The corresponding three-dimensional Euclidean action is~\cite{Linde:1980tt,Linde:1981zj}
\begin{equation}
S_3(T)
=
4\pi\int_0^\infty dr\,r^2
\left[
\frac{1}{2}
\left|\frac{d\boldsymbol{\varphi}}{dr}\right|^2
+
V_{\mathrm{eff}}(\boldsymbol{\varphi}(r),T)
-
V_{\mathrm{eff}}(\boldsymbol{\varphi}_{\mathrm{false}},T)
\right].
\label{eq:bounce_action}
\end{equation}
The bounce solution satisfies $d\boldsymbol{\varphi}/dr|_{r=0}=0$ and $\boldsymbol{\varphi}(r\to\infty)=\boldsymbol{\varphi}_{\mathrm{false}}$.
These boundary conditions ensure regularity at the bubble center and approach to the false vacuum at spatial infinity, respectively.
The corresponding bubble nucleation rate per unit time and unit volume is approximated by
\begin{equation}
\Gamma(T)
\simeq
T^4
\left(\frac{S_3(T)}{2\pi T}\right)^{3/2}
\exp\!\left[-\frac{S_3(T)}{T}\right].
\label{eq:nucleation_rate}
\end{equation}
The ratio $S_3(T)/T$ controls the exponential suppression of thermal bubble nucleation.

The nucleation and percolation temperatures are determined numerically from the resulting tunneling action under the radiation-dominated approximation, with the Hubble expansion rate given by
\begin{equation}
H(T)=1.66\sqrt{g_*}\frac{T^2}{M_{\mathrm{Pl}}},
\qquad
g_*=106.75,
\qquad
M_{\mathrm{Pl}}=1.22\times10^{19}~\mathrm{GeV}.
\label{eq:hubble_radiation}
\end{equation}
The nucleation temperature $T_n$ is defined by approximately one bubble nucleated per Hubble volume~\cite{Enqvist:1991xw,Ellis:2018mja},
\begin{equation}
N(T_n)
=
\int_{T_n}^{T_c}
\frac{dT}{T}\,
\frac{\Gamma(T)}{H^4(T)}
=1.
\label{eq:nucleation_temperature}
\end{equation}
To describe the subsequent expansion and overlap of bubbles, we introduce the false-vacuum fraction $P_{\mathrm{f}}(T)=\exp[-I(T)]$~\cite{Guth:1979bh,Guth:1981uk}, where
\begin{equation}
I(T)
=
\frac{4\pi}{3}
\int_T^{T_c}dT'\,
\frac{\Gamma(T')}{T'^4H(T')}
\left[
\int_T^{T'}d\widetilde{T}\,
\frac{v_w}{H(\widetilde{T})}
\right]^3.
\label{eq:percolation_integral}
\end{equation}
Here, $v_w$ is the bubble-wall velocity, taken to be temperature independent during a given transition.
The percolation temperature $T_p$ is determined by $I(T_p)=0.34$, corresponding to approximately $29\%$ of the volume having converted to the new phase~\cite{Ellis:2018mja}.
The physical false-vacuum volume in a fixed comoving region scales as $\mathcal{V}_{\mathrm{false}}\propto a^3(t)P_{\mathrm{f}}(t)$, where $a(t)$ is the cosmological scale factor.
All samples included in our GW analysis satisfy the condition for this volume to decrease at $T_p$~\cite{Ellis:2018mja},
\begin{equation}
3+T_p\left.\frac{dI}{dT}\right|_{T=T_p}<0.
\label{eq:false_vacuum_volume}
\end{equation}
The completion temperature $T_f$ is defined here by $I(T_f)=4.6$, at which approximately $1\%$ of the volume remains in the false vacuum.
In the numerical evaluation, the upper integration limits are replaced by the highest retained temperature in the tunneling-action grid below $T_c$.

The transition strength and characteristic inverse duration are evaluated at $T_p$.
Defining the free-energy difference between the two phases as
\begin{equation}
\Delta V(T)
=
V_{\mathrm{eff}}\!\left(\boldsymbol{\varphi}_{\mathrm{false}}(T),T\right)
-
V_{\mathrm{eff}}\!\left(\boldsymbol{\varphi}_{\mathrm{true}}(T),T\right),
\label{eq:free_energy_difference}
\end{equation}
we use the trace-anomaly definition of the strength parameter~\cite{Caprini:2019egz},
\begin{equation}
\alpha
=
\left.
\frac{\displaystyle
\Delta V(T)-\frac{T}{4}\frac{d\Delta V(T)}{dT}}
{\rho_{\mathrm{rad}}(T)}
\right|_{T=T_p},
\qquad
\rho_{\mathrm{rad}}(T)=\frac{\pi^2}{30}g_*T^4.
\label{eq:transition_strength}
\end{equation}
The inverse duration in units of the Hubble rate is obtained from
\begin{equation}
\frac{\beta}{H_p}
=
T_p
\left.
\frac{d}{dT}\left(\frac{S_3(T)}{T}\right)
\right|_{T=T_p},
\qquad
H_p\equiv H(T_p).
\label{eq:inverse_duration}
\end{equation}
These quantities, together with the bubble-wall velocity, provide the inputs for the GW calculation discussed below.

\subsection{Gravitational wave signals and detectability}
\label{subsec:gw_signals}

A first-order phase transition can generate a stochastic GW background through the dynamics of expanding bubbles and the surrounding plasma~\cite{Kamionkowski:1993fg}.
In general, the GW background receives contributions from scalar-field bubble collisions, sound waves, and magnetohydrodynamic turbulence~\cite{Caprini:2015zlo,Caprini:2019egz}.
The acoustic signal is sourced by bulk fluid motion induced by expanding bubbles~\cite{Hindmarsh:2013xza}, while turbulent plasma motions and magnetic fields can provide an additional contribution~\cite{Kosowsky:2001xp,Caprini:2009yp}.
We assume non-runaway bubble walls, for which the energy stored in the walls is small compared with that transferred to the plasma.
The scalar-field contribution is therefore neglected, and the present-day GW spectrum is approximated by
\begin{equation}
h^2\Omega_{\mathrm{GW}}(f)
\simeq
h^2\Omega_{\mathrm{sw}}(f)
+
h^2\Omega_{\mathrm{turb}}(f).
\label{eq:gw_total_spectrum}
\end{equation}
Here, $\Omega_{\mathrm{GW}}(f)=\rho_{\mathrm{c},0}^{-1}d\rho_{\mathrm{GW},0}/d\ln f$, where $\rho_{\mathrm{c},0}$ is the present critical density, and $h=H_0/(100~\mathrm{km\,s^{-1}\,Mpc^{-1}})$.

The bubble-wall velocity is estimated using the phenomenological Chapman--Jouguet prescription~\cite{Steinhardt:1981ct,Espinosa:2010hh},
\begin{equation}
v_{\mathrm{CJ}}(\alpha)
=
\frac{1/\sqrt{3}+\sqrt{\alpha^2+2\alpha/3}}{1+\alpha}.
\label{eq:cj_wall_velocity}
\end{equation}
Numerically, the wall velocity is fixed using a preliminary percolation solution and is then held constant when updating $T_p$ and the transition parameters.
For the conversion of released energy into bulk fluid motion, we use the Chapman--Jouguet efficiency fit~\cite{Espinosa:2010hh}.
Assuming that $10\%$ of the bulk kinetic energy is converted into turbulence~\cite{Caprini:2015zlo}, the efficiency factors are given by
\begin{equation}
\kappa_v
=
\frac{\sqrt{\alpha}}{0.135+\sqrt{0.98+\alpha}},
\qquad
\kappa_{\mathrm{turb}}=0.1\,\kappa_v.
\label{eq:gw_efficiency_factors}
\end{equation}
Throughout the following expressions, $\alpha$ is the trace-anomaly parameter evaluated at $T_p$, and we introduce $\widetilde{\beta}\equiv\beta/H_p$.

To account approximately for reheating, the temperature entering the redshift factors is taken to be
\begin{equation}
T_\star
=
T_p\left[1+\max(\alpha_{\mathrm{e}},0)\right]^{1/4},
\qquad
\alpha_{\mathrm{e}}
=
\left.
\frac{\displaystyle\Delta V-T\frac{d\Delta V}{dT}}
{\rho_{\mathrm{rad}}(T)}
\right|_{T=T_p}.
\label{eq:gw_reheating_temperature}
\end{equation}
Here, $\alpha_{\mathrm{e}}$ is the energy-density difference between the two phases normalized to the radiation energy density at $T_p$ and is distinct from the trace-anomaly parameter $\alpha$.
The transition parameters are evaluated at percolation, whereas $T_\star$ is used for redshifting the GW spectrum to the present epoch.

With these conventions, the sound-wave and turbulence spectra are written as~\cite{Hindmarsh:2015qta,Caprini:2015zlo}
\begin{equation}
\begin{aligned}
h^2\Omega_{\mathrm{sw}}(f)
={}&
2.65\times10^{-6}
\frac{v_w}{\widetilde{\beta}}
\left(\frac{\kappa_v\alpha}{1+\alpha}\right)^2
\left(\frac{100}{g_*}\right)^{1/3}
\Upsilon_{\mathrm{sw}}\,S_{\mathrm{sw}}(f),
\\
h^2\Omega_{\mathrm{turb}}(f)
={}&
3.35\times10^{-4}
\frac{v_w}{\widetilde{\beta}}
\left(\frac{\kappa_{\mathrm{turb}}\alpha}{1+\alpha}\right)^{3/2}
\left(\frac{100}{g_*}\right)^{1/3}
S_{\mathrm{turb}}(f),
\end{aligned}
\label{eq:gw_component_spectra}
\end{equation}
where the spectral shapes are
\begin{equation}
\begin{aligned}
S_{\mathrm{sw}}(f)
&=
\left(\frac{f}{f_{\mathrm{sw}}}\right)^3
\left[
\frac{7}{4+3(f/f_{\mathrm{sw}})^2}
\right]^{7/2},
\\
S_{\mathrm{turb}}(f)
&=
\frac{(f/f_{\mathrm{turb}})^3}
{\left(1+f/f_{\mathrm{turb}}\right)^{11/3}
 \left(1+8\pi f/h_*\right)}.
\end{aligned}
\label{eq:gw_spectral_shapes}
\end{equation}
The characteristic frequencies redshifted to today are
\begin{equation}
\begin{aligned}
f_{\mathrm{sw}}
&=
1.9\times10^{-5}~\mathrm{Hz}\,
\frac{\widetilde{\beta}}{v_w}
\left(\frac{T_\star}{100~\mathrm{GeV}}\right)
\left(\frac{g_*}{100}\right)^{1/6},
\\
f_{\mathrm{turb}}
&=
2.7\times10^{-5}~\mathrm{Hz}\,
\frac{\widetilde{\beta}}{v_w}
\left(\frac{T_\star}{100~\mathrm{GeV}}\right)
\left(\frac{g_*}{100}\right)^{1/6},
\\
h_*
&=
1.65\times10^{-5}~\mathrm{Hz}\,
\left(\frac{T_\star}{100~\mathrm{GeV}}\right)
\left(\frac{g_*}{100}\right)^{1/6}.
\end{aligned}
\label{eq:gw_characteristic_frequencies}
\end{equation}
Here, $h_*$ denotes the redshifted Hubble frequency.

The finite lifetime of the acoustic source is included through the suppression factor~\cite{Ellis:2020awk,Guo:2020grp}
\begin{equation}
\Upsilon_{\mathrm{sw}}
=
1-\frac{1}{\sqrt{1+2H_p\tau_{\mathrm{sw}}}},
\qquad
\tau_{\mathrm{sw}}\simeq\frac{R_*}{\overline{U}_f},
\label{eq:sound_wave_lifetime}
\end{equation}
with the characteristic bubble separation and enthalpy-weighted root-mean-square fluid velocity approximated by
\begin{equation}
H_pR_*
=
(8\pi)^{1/3}\frac{v_w}{\widetilde{\beta}},
\qquad
\overline{U}_f^{\,2}
\simeq
\frac{3\kappa_v\alpha}{4(1+\alpha)}.
\label{eq:bubble_separation_fluid_velocity}
\end{equation}
The peak frequency $f_{\mathrm{peak}}$ and peak amplitude $h^2\Omega_{\mathrm{GW}}^{\mathrm{peak}}\equiv h^2\Omega_{\mathrm{GW}}(f_{\mathrm{peak}})$ are extracted numerically from the total spectrum in Eq.~\eqref{eq:gw_total_spectrum}.

We consider the detection prospects for LISA~\cite{LISA:2017pwj}, TianQin~\cite{TianQin:2015yph,Hu:2018yqb,TianQin:2020hid}, DECIGO~\cite{Seto:2001qf,Kawamura:2020pcg}, and BBO~\cite{Crowder:2005nr,Corbin:2005ny}.
The signal-to-noise ratio (SNR) is calculated by integrating the full spectrum over each detector's frequency range~\cite{Kudoh:2005as,Schmitz:2020syl},
\begin{equation}
\mathrm{SNR}
=
\left[
n_{\mathrm{det}}\,\mathcal{T}
\int_{f_{\min}}^{f_{\max}}df\,
\left(
\frac{h^2\Omega_{\mathrm{GW}}(f)}
     {h^2\Omega_{\mathrm{noise}}(f)}
\right)^2
\right]^{1/2}.
\label{eq:gw_snr}
\end{equation}
Here, $\Omega_{\mathrm{noise}}(f)$ is the effective noise spectrum expressed in GW energy-density units~\cite{Moore:2014lga}, and $\mathcal{T}$ denotes the effective observation time.
Following the adopted noise-curve convention, we use $n_{\mathrm{det}}=1$ for LISA and TianQin and $n_{\mathrm{det}}=2$ for DECIGO and BBO, corresponding to auto-correlation and cross-correlation searches, respectively.
The LISA, DECIGO, and BBO noise curves are taken from Ref.~\cite{Schmitz:2020syl}, while the TianQin noise model uses the design specifications in Ref.~\cite{TianQin:2020hid}.
The adopted LISA sensitivity assumes perfectly characterized instrumental noise and neglects astrophysical foregrounds, including confusion noise from unresolved white-dwarf binaries~\cite{Schmitz:2020syl}.
We adopt $\mathcal{T}=3~\mathrm{yr}$ and classify signals with a full-spectrum $\mathrm{SNR}\geq10$ as potentially detectable.
For visual comparison, we also construct power-law-integrated (PLI) sensitivity curves with the same observation time and SNR threshold~\cite{Thrane:2013oya}.

\subsection{Numerical results and discussion}
\label{subsec:pt_gw_results}

In the numerical calculation, the bounce action $S_3(T)/T$ is evaluated on a nonuniform temperature grid, with denser sampling around the preliminary nucleation temperature, and interpolated using a shape-preserving piecewise cubic Hermite interpolant.
The temperature derivative entering $\alpha$ is evaluated using a central difference with $\Delta T_\alpha=\max(0.02~\mathrm{GeV},10^{-3}T)$.
For $\beta/H_p$, we compare central-difference estimates around $T_p$ using $\Delta T_\beta=0.15$, $0.30$, $0.60$, and $1.20~\mathrm{GeV}$, supplemented by a local cubic fit within $|T-T_p|\leq2~\mathrm{GeV}$ when sufficient action samples are available.
The adopted value is the median of the available finite positive estimates obtained with the two larger steps and the local fit, while the derivative of the interpolant is used as a fallback.

\begin{figure*}[!ht]
    \centering
    \setlength{\abovecaptionskip}{-5pt}
    \includegraphics[width=\linewidth]{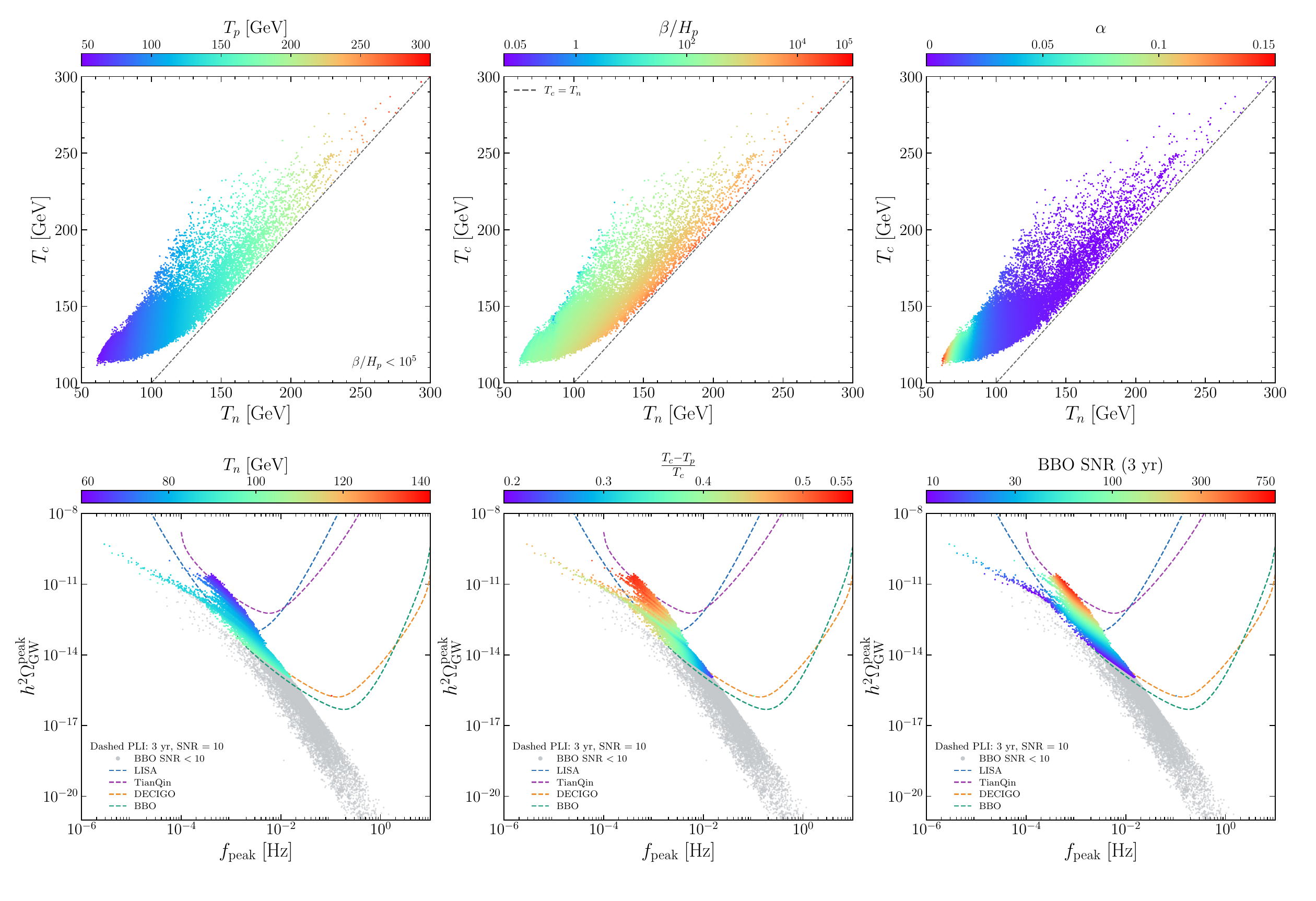}
    \caption{Phase-transition parameters and GW peak properties of the selected samples.
The upper panels show the samples in the $T_n$--$T_c$ plane, with colors indicating $T_p$, $\beta/H_p$, and $\alpha$ from left to right.
Only samples with $\beta/H_p<10^5$ are displayed in the upper panels, where the gray dashed lines denote $T_c=T_n$.
The lower panels show the peak amplitude $h^2\Omega_{\mathrm{GW}}^{\mathrm{peak}}$ versus the peak frequency $f_{\mathrm{peak}}$, with colors indicating $T_n$, $(T_c-T_p)/T_c$, and the BBO SNR from left to right.
Colored samples satisfy $\mathrm{SNR}_{\mathrm{BBO}}\geq10$, whereas samples below this threshold are shown in light gray.
The dashed sensitivity curves correspond to LISA, TianQin, DECIGO, and BBO for an observation time of three years and an SNR threshold of $10$.
They are constructed following Ref.~\cite{Thrane:2013oya}, using the noise models in Refs.~\cite{Schmitz:2020syl,TianQin:2020hid}.}
    \label{fig:pt_gw_results}
\end{figure*}

The samples used in the following analysis satisfy the nucleation, percolation, and transition-completion criteria, and their GW spectra are evaluated numerically.
Figure~\ref{fig:pt_gw_results} summarizes the resulting phase-transition properties and GW predictions.
The upper panels display the samples in the $T_n$--$T_c$ plane, with the colors indicating $T_p$, $\beta/H_p$, and $\alpha$ from left to right.
The dashed diagonal denotes $T_c=T_n$, and only samples with $\beta/H_p<10^5$ are displayed in these panels.
The lower panels show the peak amplitude $h^2\Omega_{\mathrm{GW}}(f_{\mathrm{peak}})$ as a function of the peak frequency $f_{\mathrm{peak}}$, with the colors representing $T_n$, the relative supercooling $(T_c-T_p)/T_c$, and the BBO SNR, respectively.
Here, colored samples satisfy $\mathrm{SNR}_{\mathrm{BBO}}\geq10$, whereas samples below this threshold are shown in gray.
The dashed curves show the PLI sensitivity curves~\cite{Thrane:2013oya} of LISA, TianQin, DECIGO, and BBO, constructed for an observation time of three years and an SNR threshold of 10.

Several features can be inferred from Fig.~\ref{fig:pt_gw_results}:
\begin{itemize}
\item
As shown in the upper-left panel, the critical temperature $T_c$ can be as low as approximately $112~\mathrm{GeV}$, while the nucleation and percolation temperatures reach approximately $61~\mathrm{GeV}$ and $54~\mathrm{GeV}$, respectively.
For these low-temperature transitions, nucleation and percolation occur well below the critical temperature, indicating appreciable supercooling.

\item
The upper-middle panel shows that, for comparable values of $T_c$, a larger separation between $T_c$ and $T_n$ tends to be associated with a smaller $\beta/H_p$, which can fall to $\mathcal{O}(10)$ or below.
By contrast, samples close to the diagonal $T_c=T_n$ generally exhibit larger $\beta/H_p$, approaching $10^5$ within the displayed range.
Meanwhile, the upper-right panel shows that larger values of $\alpha$, reaching approximately $0.15$, preferentially occur in the low-temperature region with appreciable supercooling and small $T_p$. 

\item
The lower panels exhibit an overall anticorrelation between the peak frequency and the peak amplitude, with higher-frequency signals generally having smaller amplitudes.
Using the full-spectrum SNR criterion for three years of observation, the samples accessible to LISA are predominantly concentrated around $f_{\mathrm{peak}}\sim10^{-3}~\mathrm{Hz}$, whereas a smaller subset near $f_{\mathrm{peak}}\sim5\times10^{-4}~\mathrm{Hz}$ also reaches $\mathrm{SNR}\geq10$ for TianQin.
Meanwhile, DECIGO and BBO probe a larger fraction of the displayed samples, with detectable signals extending to peak frequencies of approximately $8\times10^{-3}~\mathrm{Hz}$ and $10^{-1}~\mathrm{Hz}$, respectively.

\item
At comparable peak frequencies, the lower-left panel shows that smaller nucleation temperatures $T_n$ tend to be associated with larger peak amplitudes.
Similarly, the lower-middle panel indicates that samples with larger $(T_c-T_p)/T_c$, corresponding to more pronounced supercooling, preferentially yield higher peak amplitudes.
Finally, the lower-right panel shows that the BBO SNR can reach approximately $750$ for three years of observation, with the largest value occurring at $f_{\mathrm{peak}}\simeq7.2\times10^{-4}~\mathrm{Hz}$.

\end{itemize}

\begin{table*}[!ht]
\centering
\caption{Phase-transition parameters and GW observables for the benchmark points listed in Table~\ref{tab:benchmark_points}.
The SNRs assume three years of observation.}
\label{tab:benchmark_gw}
\begingroup
\small
\setlength{\tabcolsep}{5pt}
\renewcommand{\arraystretch}{1.05}
\begin{tabular}{@{}lccccc@{}}
\toprule
 & Point 1 & Point 2 & Point 3 & Point 4 & Point 5 \\
\midrule
$T_c\ [\mathrm{GeV}]$ & 127.67 & 135.73 & 138.08 & 117.85 & 114.10 \\
$T_n\ [\mathrm{GeV}]$ & 76.31 & 87.63 & 90.99 & 64.23 & 62.06 \\
$T_p\ [\mathrm{GeV}]$ & 67.62 & 80.80 & 84.87 & 57.42 & 55.44 \\
$T_\star\ [\mathrm{GeV}]$ & 68.87 & 81.59 & 85.61 & 59.34 & 57.57 \\
$T_f\ [\mathrm{GeV}]$ & 65.30 & 78.73 & 83.25 & 54.93 & 53.19 \\
$\alpha(T_p)$ & 0.05736 & 0.02757 & 0.02334 & 0.10999 & 0.12742 \\
$\beta/H_p$ & 56.21 & 65.26 & 110.45 & 39.73 & 42.22 \\
$v_w$ & 0.7356 & 0.6947 & 0.6867 & 0.7802 & 0.7913 \\
$f_{\mathrm{peak}}\ [\mathrm{mHz}]$ & 1.0116 & 1.4750 & 2.6461 & 0.5805 & 0.5902 \\
$h^2\Omega_{\mathrm{GW}}^{\mathrm{peak}}\ [10^{-12}]$ & 1.1298 & 0.1349 & 0.0374 & 10.0678 & 12.9580 \\
$\mathrm{SNR}_{\mathrm{LISA}}$ & 46.10 & 8.43 & 3.21 & 177.80 & 235.18 \\
$\mathrm{SNR}_{\mathrm{TianQin}}$ & 2.59 & 0.54 & 0.31 & 9.73 & 12.84 \\
$\mathrm{SNR}_{\mathrm{DECIGO}}$ & 80.14 & 17.74 & 11.91 & 218.04 & 292.34 \\
$\mathrm{SNR}_{\mathrm{BBO}}$ & 172.65 & 38.22 & 25.66 & 469.80 & 629.86 \\
\bottomrule
\end{tabular}
\endgroup
\end{table*}

\begin{figure*}[!ht]
    \centering
    \setlength{\abovecaptionskip}{-5pt}
    \includegraphics[width=\linewidth]{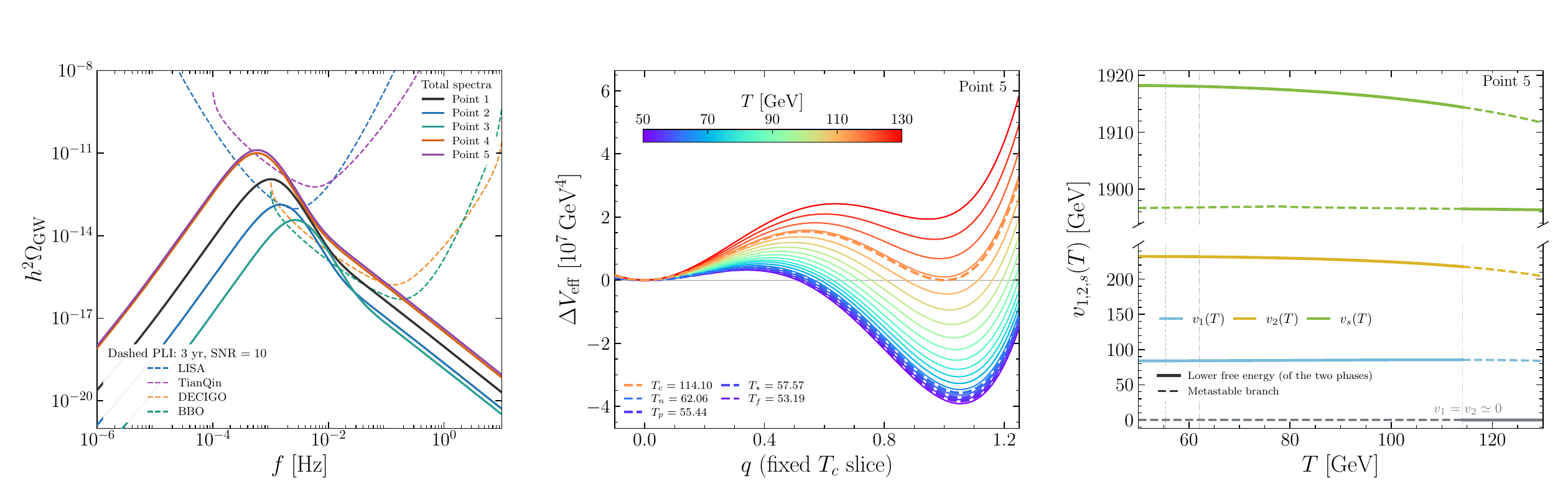}
    \caption{GW spectra and phase-transition dynamics of representative benchmark points.
Left: total GW spectra of the five benchmark points listed in Tables~\ref{tab:benchmark_points} and \ref{tab:benchmark_gw}.
The dashed sensitivity curves correspond to LISA, TianQin, DECIGO, and BBO for an observation time of three years and an SNR threshold of $10$.
They are constructed following Ref.~\cite{Thrane:2013oya}, using the noise models in Refs.~\cite{Schmitz:2020syl,TianQin:2020hid}.
Middle: effective-potential profiles for Point 5 at different temperatures along the same straight-line slice connecting the false and true vacua at $T_c$, corresponding to $q=0$ and $q=1$, respectively.
The potential difference $\Delta V_{\mathrm{eff}}$ is measured relative to its value at $q=0$ at each temperature.
Colors indicate the temperature, and dashed profiles highlight $T_c$, $T_n$, $T_p$, $T_\star$, and $T_f$.
Right: temperature dependence of $v_1$, $v_2$, and $v_s$ for Point 5, shown in light blue, gold, and green, respectively, with a broken vertical axis.
Solid and dashed curves show the lower-free-energy and metastable branches, respectively, of the two phases involved in the transition.
The vertical lines mark $T_c$, $T_n$, and $T_p$.}
    \label{fig:benchmark_gw}
\end{figure*}

The five benchmark points presented in the previous section are used to examine the dynamics of the completed first-order phase transitions and their associated GW signals.
Their total GW spectra are shown in the left panel of Fig.~\ref{fig:benchmark_gw}.
For three years of observation and an SNR threshold of $10$, all five benchmark points are potentially detectable at BBO and DECIGO.
The BBO SNRs range from $25.66$ for Point 3 to $629.86$ for Point 5, which also yields the largest DECIGO SNR of $292.34$.
Points 1, 4, and 5 exceed this threshold at LISA, and Point 5 exceeds it at TianQin.
Point 5 yields the largest SNRs at LISA and TianQin, $235.18$ and $12.84$, respectively.
The effective-potential profiles and temperature-dependent scalar VEVs for Point 5 are shown in the middle and right panels, respectively.

The GW-producing transitions of these five points share the same electroweak-symmetry-breaking pattern,
\begin{equation}
(0,0,v_s^{\mathrm{false}})
\longrightarrow
(v_1^{\mathrm{true}},v_2^{\mathrm{true}},v_s^{\mathrm{true}}),
\qquad
v_s^{\mathrm{true}}>v_s^{\mathrm{false}}>0,
\label{eq:benchmark_transition_pattern}
\end{equation}
where the field values on both sides are evaluated at the same temperature.
Thus, both Higgs doublets acquire nonzero VEVs in the same first-order transition, whereas the singlet already has a nonzero expectation value in the false vacuum and undergoes a further discontinuous increase.

For all five benchmark points, we cross-check the tunneling actions at $T_p$ and $T_p\pm\Delta T$, with $\Delta T=0.3$, $0.6$, and $1.2~\mathrm{GeV}$, using the independent bounce solver of \texttt{BSMPT-v3}~\cite{Basler:2018cwe,Basler:2020nrq,Basler:2024aaf}, with the same finite-temperature effective potential as in the \texttt{CosmoTransitions} calculation.
The relative differences in $S_3(T)/T$ remain below $0.30\%$.
For $\beta/H_p$, we compare the \texttt{BSMPT-v3} results with the original \texttt{CosmoTransitions} estimates using the same finite-difference step size for each benchmark point.
The largest relative difference is $9.88\%$ for Point 1, while the differences for the other four points are below $5\%$, with the smallest being $1.59\%$ for Point 3.
An independent minimization of the same effective potential yields $\alpha(T_p)$ with relative differences below $2.4\times10^{-6}$.

\section{Conclusions}
\label{sec:conclusions}

In this work, we have investigated the Type-I S2HDM, whose CP-odd singlet component serves as a pNGB DM candidate.
Identifying $h_1$ with the observed $125~\mathrm{GeV}$ Higgs boson and considering $h_2$ near $152~\mathrm{GeV}$, we have explored a possible connection between the $\gamma\gamma+\mathrm{MET}$ excesses, the DM relic density, and GWs from a first-order electroweak phase transition.
The parameter space has been investigated through random scans supplemented by MCMC sampling, taking into account the relevant theoretical and experimental constraints.

The $\gamma\gamma+\mathrm{MET}$ signature arises from the cascade $gg\to h_3\to h_2h_2\to\gamma\gamma\chi\chi$.
Our collider simulations show that the mass of $h_3$ plays an important role in determining the MET distribution: increasing $m_{h_3}$ generally produces a harder spectrum and favors the High-MET category, whereas lighter $h_3$ states preferentially populate the Low-MET category.
The model admits parameter regions compatible with either excess individually and an overlapping region compatible with both within their respective $1\sigma$ intervals.
The overlapping region also contains samples satisfying the relic-density requirement.

We have further examined the finite-temperature phase-transition dynamics and the associated GW signals for samples that explain at least one of the two $\gamma\gamma+\mathrm{MET}$ excesses within $1\sigma$ and reproduce the observed DM relic density.
The stronger GW signals are associated with appreciable supercooling, larger transition strengths, and relatively small values of $\beta/H_p$.
For three years of observation and an SNR threshold of $10$, parts of the predicted signal population are accessible to LISA, DECIGO, and BBO, while some of the stronger signals also reach the sensitivity of TianQin.

In particular, Point 1 simultaneously accommodates the $\gamma\gamma+\mathrm{MET}$ excesses in the Low-MET and High-MET regions within their respective $1\sigma$ intervals and yields $\Omega_\chi h^2=0.11635$, consistent with the observed DM relic density within the assumed $10\%$ theoretical uncertainty.
Its first-order electroweak phase transition produces a GW signal with SNRs of approximately $46.10$, $2.59$, $80.14$, and $172.65$ at LISA, TianQin, DECIGO, and BBO, respectively, for three years of observation.
These results highlight the complementarity of collider searches, DM measurements, and future space-based GW observations in probing the S2HDM.

\acknowledgments
We thank Yifan Zhu for helpful discussions. 
This work was supported by the National Natural Science Foundation of China under Grant No. 12275066 and by the startup research funds of Henan University. 
The work of H. Yang was also supported by the National Natural Science Foundation of China under Grant No. W2441004.
The work of K. Wang was supported by the Open Project of the Shanghai Key Laboratory for Particle Physics and Cosmology under Grant No. 22DZ2229013-3.

\appendix


\bibliographystyle{apsrev4-1}
\bibliography{apssamp}

\end{document}